\documentclass[aps,reprint,nofootinbib,nobibnotes,notitlepage,superscriptaddress,twocolumn,prd,amsmath,amssymb
]{revtex4-2}

\usepackage[T1]{fontenc}
\usepackage{aecompl}
\usepackage{natbib}
\usepackage{verbatim}
\usepackage[usenames,dvipsnames]{xcolor}
\usepackage{tabulary}
\usepackage{tabularx}
\usepackage{todonotes}
\usepackage{hyperref}
\hypersetup{
	colorlinks = true,
    linkcolor = blue,
    urlcolor  = blue,
    citecolor = blue
}
\usepackage{braket}
\usepackage{float}
\usepackage{dcolumn}
\usepackage{bm}
\usepackage{microtype}
\usepackage{wasysym}

\newcommand{\aap}{A\&A}

\newcommand{\aj}{A.J.}
\newcommand{\mnras}{MNRAS}

\newcommand{\abacussummit}{{\tt AbacusSummit}}
\newcommand{\abacushod}{{\tt AbacusHOD}}

\numberwithin{equation}{section}

\defcitealias{Asgari2021}{A21}
\defcitealias{Hildebrandt2021}{H21}
\defcitealias{Takahashi2017}{T17}
\defcitealias{Heydenreich2023}{H23}
\defcitealias{Planck2020}{PL18}

\newcommand{\ttheta}[1]{\pmb{\theta_\mathrm{ap}}}

\newcommand{\Norm}[2]{\mathcal{N}}

\newcommand{\Mpch}{\,h^{-1}\text{Mpc}}
\newcommand{\Gpch}{\,h^{-1}\text{Gpc}}
\newcommand{\Omegam}{\Omega_{\rm m}}

\newcommand{\omegac}{\omega_{\rm cdm}}
\newcommand{\omegab}{\omega_{\rm b}}

\newcommand{\Neff}{N_{\rm eff}}
\newcommand{\Msunh}{\,h^{-1}{\rm M_{\odot}}}

\newcommand{\nrun}{{\rm d}\, n_\mathrm{s}/{\rm d}\ln k}
    
\newcommand{\unif}{\mathcal{U}}

\newcommand\be{\begin{equation}}
\newcommand\ee{\end{equation}}
\def\bea{\begin{eqnarray}}
\def\eea{\end{eqnarray}}

\allowdisplaybreaks
\begin{document}

\title{Cosmological parameters from the joint analysis of Density Split and Second Order Statistics: an Emulator based on the Halo Occupation Distribution}

\author{Pierre A. Burger}
\email[Corresponding author: ]{pierre.burger@uwaterloo.ca}
\affiliation{Waterloo Centre for Astrophysics, University of Waterloo, Waterloo, ON N2L 3G1, Canada}
\affiliation{Department of Physics and Astronomy, University of Waterloo, Waterloo, ON N2L 3G1, Canada}
\author{Enrique Paillas}
\affiliation{Waterloo Centre for Astrophysics, University of Waterloo, Waterloo, ON N2L 3G1, Canada}
\affiliation{Department of Physics and Astronomy, University of Waterloo, Waterloo, ON N2L 3G1, Canada}
\author{Michael J. Hudson}
\affiliation{Department of Physics and Astronomy, University of Waterloo, Waterloo, ON N2L 3G1, Canada}
\affiliation{Waterloo Centre for Astrophysics, University of Waterloo, Waterloo, ON N2L 3G1, Canada}
\affiliation{Perimeter Institute for Theoretical Physics, Waterloo, ON N2L 2Y5, Canada}


\date{\today}


\label{firstpage}
\begin{abstract}
In this work, we develop a simulation-based model to predict the density split (DSS) and second-order shear and clustering statistics. A simulation-based model has the potential to model highly non-linear scales where current DSS models fail. To build this model, we use the \texttt{AbacusSummit} N-body simulation suite from which we measure all necessary statistics and train an emulator based on \texttt{CosmoPower}.  In that context, we discuss possible improvements for future emulators to make the measurement less noisy and biased, resulting in more accurate and precise model predictions.

Regarding the emulator's accuracy, we find that the most important aspect is the average of the summary statistics over multiple-shot noise realizations of the foreground galaxies. However, these results probably depend on the chosen number density of the foreground galaxies. Regarding the parameter forecast based on preliminary LOWZxUNIONS data, we find that DSS has more constraining power to derive cosmological parameters than second-order statistics and that the joint analysis with second-order statistics is particularly useful for extracting parameters of the galaxy-halo connection.
\end{abstract}

\pacs{Valid PACS appear here}

\maketitle

\section{Introduction}
\label{sec:introduction}

To infer cosmological parameters of the standard model of cosmology, called the $\Lambda$ Cold Dark Matter model ($\Lambda$CDM), most analyses focus on second-order statistics like the two-point correlation functions and its Fourier counterpart, the power spectrum. This is due to accurate theoretical predictions and the control over systematic inaccuracies. These second-order statistics can extract all the Gaussian information contained in the data. In the late Universe, however, non-linear gravitational instabilities generate many non-Gaussian features whose information can only be accessed with higher-order statistics. Furthermore, since higher-order statistics scale differently with cosmology, the constraining power on cosmological parameters increases by jointly investigating second- and higher-order statistics \citep{HOWLS2023}. Recently used examples of higher-order statistics that make use of gravitational lensing, which describes the deflection of light by massive objects (shear), are the peak count statistics \citep{Martinet:2018,Harnois-Deraps:2021,Harnois2024}, persistent homology \citep{Heydenreich:2021,Heydenreich2022}, and the integrated three-point correlation function used in \citep[][]{Halder:2021,Halder:2022}, along with a second- and third-order shear analysis \citep{Gatti2022,Burger2024}. Although the use of galaxy clustering data introduces new unknown astrophysical parameters like the galaxy bias, which describes the connection of the total matter and galaxies, it has been shown that by jointly investigating weak lensing and galaxy clustering data \citep{vanUitert2018,Joudaki2018,DES:2018, DES2021} constraints on $\sigma_8$ and $\Omega_\mathrm{m}$ are improved, where $\sigma_8$ is the amplitude of matter fluctuations in spheres of $8 \Mpch$, and $\Omegam$ is the matter density parameter. 

To extract the additional information in the non-Gaussian features, \cite{Gruen:2018,Burger2023,Paillas2023} made use of the density-split statistics (DSS), the general idea of which is to define environments through a smoothed version of the foreground galaxy distribution, and then correlate these environments with either the positions or shear of galaxies. The former we refer to as density split clustering (DSC) and the latter as density split lensing (DSL). In \cite{Gruen:2018} and \cite{Burger2023}, an analytical model \citep{Friedrich:2018, Burger2022} of the DSL was used to infer cosmological and galaxy-halo parameters like the galaxy bias, and both works showed that the DSL is competitive with recent second-order analysis. A recent application of the DSC to three-dimensional clustering of galaxies from the Baryon Oscillation Spectroscopic Survey \citep[BOSS;][]{Dawson2013} was made in \cite{Paillas2023}. To derive cosmological parameters, they used a simulation-based model \citep{Cuesta-Lazaro2023} based on the \texttt{AbacusSummit} simulations \citep{Maksimova2021}.

Analytical models have the advantage of better dissecting the basic underlying properties of the large-scale structure (LSS), and they can compute the summary statistics noise-free at any point in the parameter space. However, the DSL analytical models \cite{Friedrich:2018, Burger2023} rely on several assumptions, like the galaxy bias, that are only valid if large scales are considered. More complicated non-linear galaxy bias models must be developed to model the galaxy-matter connection on small scales or use a halo model approach. A halo model assumes that all matter is contained in spherical haloes and that all galaxies reside in these haloes \cite{Seljak2000,Cooray2002}.

Although simulation-based models suffer from noisy measured summary statistics and depend on the number of training nodes for the accuracy of the interpolation to any point in the parameter space, they have the advantage that smaller scales can be considered without the need for complicated analytical descriptions. Based on the halo model, the galaxy-matter connection can be simulated using a Halo Occupation Distribution \cite[HOD;][]{Peacock2000}, which describes a method to distribute galaxies into haloes based on properties like mass of the halo.

For this work, we follow the idea of \cite{Cuesta-Lazaro2023} and build a model based on the \texttt{AbacusSummit}\footnote{\url{https://abacussummit.readthedocs.io}} simulations. For all the \texttt{AbacusSummit} simulations with their different cosmologies and HOD parameters, we measure our summary statistics and use an emulator approach to predict the model everywhere in the predefined parameters space. Emulators are a machine learning tool widely used in cosmology due to their speed and accuracy in interpolating 
the model vector between predefined training nodes \cite{COSMOPOWER2022,EE2021,Angulo2021,Bonici2024}.
Furthermore, we combine the works of \cite{Paillas2023} and \cite{Gruen:2018}. This means that we build a model that predicts the angular clustering of the foreground galaxies $w(\theta)$, the galaxy-galaxy lensing $g_\mathrm{t}(\theta)$, DSL, and the correlation between the foreground environments with the galaxies that were used to define these environments in the first place, which we call density split clustering (DSC). Compared to \cite{Paillas2023}, the difference is that we restrict our analysis to the projected sky and measure all statistics in angular scales.

This work is structured as follows. In Sect.~\ref{Aperture_stat}, we review the basics of the DSS. In Sect.~\ref{sec:data}, we briefly describe the reference data used to create a realistic covariance matrix for a parameter forecast and emulator accuracy validation. In Sect.~\ref{sec:AbacusSummit} review the basics of the \texttt{AbacusSummit} simulations that we use to build our model, and in Sect.~\ref{sec:T17}, we describe estimating a numerical covariance matrix. In Sect.~\ref{sec:model_measurement}, we continue measuring the model vectors and describe in Sect.~\ref{sec:emulator} how we build our emulator model from these measurements. In Sect.~\ref{sec:forecast}, we perform a forecast and finalize our work in Sect.~\ref{sec:conclusion}.

\section{Density-split statistics}
\label{Aperture_stat}

The DSS fundamentally assesses the tangential shear $\gamma_{\mathrm{t}}(\theta)$ \citep{Schneider:1996} or angular clustering $w(\theta)$ \citep{Landy1993} surrounding sub-areas (quantiles) of the sky, allocated based on the density of foreground galaxies. Given that the matter or galaxy distribution is highly skewed, the DSS captures this non-linear information by considering all second-order statistics around those subareas, providing additional constraining information. To define these quantiles $\mathcal{Q}$, we smooth the foreground galaxy number counts $n(\boldsymbol{\theta})$ with a filter function $U$, which, motivated by \cite{Schneider:1998}, we call the aperture number
\begin{equation}
    N_{ \mathrm{ap}}(\boldsymbol{\theta}) = \int\mathrm{d}^2\theta' \,n(\boldsymbol{\theta}+\boldsymbol{\theta}')\,U(|\boldsymbol{\theta}'|)\, .
    \label{Nap}
\end{equation}
For a top-hat filter, this definition aligns with the 'Counts-in-Cell' (CiC) statistics discussed in \cite{Gruen:2015}.

The general idea of the DSS is to divide the survey area into quantiles $N_\mathcal{Q}$ according to the aperture number $N_\mathrm{ap}$ and then correlate points of each quantile with the shear of the background galaxies, resulting in $N_\mathcal{Q}$ DSL or DSC signals. Following \cite{Burger2022}, 
we measure the aperture number by distributing the foreground (lens) galaxies onto a \texttt{HEALPix} \citep{HEALPix2005} grid $n(\boldsymbol{\theta})$ of $\texttt{nside}=4096$, which resulted in a pixel area of $A\approx 0.74\,\mathrm{arcmin}^2$. Second, we determined the aperture number field $N_\mathrm{ap}$ with a $10'$ Gaussian filter function $U$. The motivation for using a Gaussian filter over a top-hat filter is that it allows for the probing of smaller scales for the same number density of galaxies. Furthermore, the Gaussian filter is less likely to have empty apertures as it extends to larger scales. The filter size was chosen so that the density PDF was smooth and did not sharply peak around 0, which could happen if apertures were empty due to shot noise. This would hinder our ability to capture density variations in the low-density tail of the PDF, making it difficult to properly define quantiles and erasing some of the cosmological information encoded in these statistics.
The smoothing was achieved with the \texttt{healpy}\footnote{\url{https://healpy.readthedocs.io}} function \texttt{smoothing}, with a beam window function that is the $U$-filter in the spherical harmonic space determined with \texttt{healpy} function \texttt{beam$2$bl}. In the presence of a masked survey, the aperture number calculation has to be modified to
\begin{align}
N_\mathrm{ap}(\boldsymbol{\theta}) =  \frac{\int_0^{\theta_\mathrm{max}} n(\boldsymbol{\theta}+\boldsymbol{\theta}')U(\boldsymbol{\theta}') \,\mathrm{d}^2 \theta' }{ \int_0^{\theta_\mathrm{max}} R(\boldsymbol{\theta}+\boldsymbol{\theta}')U(\boldsymbol{\theta}') \,\mathrm{d}^2 \theta'}
\label{eq:Nap_corrected}
\end{align}
where in this work, $R(\boldsymbol{\theta})$ is the distribution of randoms inside the mask, and $\theta_\mathrm{max}$ is a large value where $U(\theta_\mathrm{max}) \rightarrow 0$.

\begin{figure}
\includegraphics[width=\linewidth]{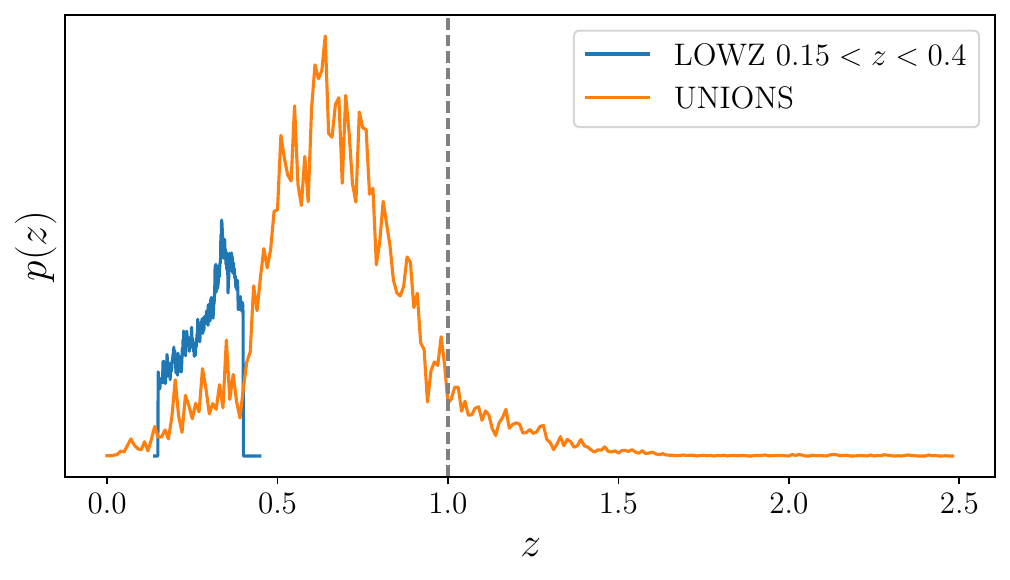}
\caption{Probability density distribution
of the redshift of the overlapping foreground LOWZ galaxies for $0.15<z<0.4$ and a preliminary UNIONS-like distribution. The dashed line indicates where we cut the model's $p(z)$ due to the light cone geometry.}
\label{fig:nz}
\end{figure}

\section{Reference data}
\label{sec:data}
For this work, we focus on developing the model and do not present any cosmological results from real data. However, in the future, we aim to apply this method to a LOWZxUNIONS setup. LOWZ is the low redshift sample of BOSS, which contains in the 12th data release spectroscopic redshift of 309527 galaxies inside an effective area of $8337\,\mathrm{deg}^2$. The Ultraviolet Near Infrared Optical Northern Survey (UNIONS) is a weak lensing survey that measures galaxy shapes and photometric redshifts North of the declination $+30^\circ$ and a galactic latitude $|b|>25^\circ$.
To perform a parameter forecast and evaluate our emulator's accuracy compared to the expected uncertainty of LOWZxUNIONS, we build a LOWZxUNIONS-like covariance matrix using a preliminary UNIONS-like\footnote{\url{https://www.skysurvey.cc/}} probability density distribution
of the redshift $p(z)$ shown in Fig.~\ref{fig:nz} and shape catalogue. The used UNIONS-like $p(z)$, estimated from overlapping spectroscopic data, and the shape catalogue are preliminary products and do not have corresponding references. Still, we refer to \cite{Guinot2022} for a description of a UNIONS shape catalogue and to \cite{Li2024} for an updated non-tomographic UNIONS $p(z)$. For a detailed description of LOWZ data, we refer to \cite{Reid2016}.

\begin{figure}
\includegraphics[width=\linewidth]{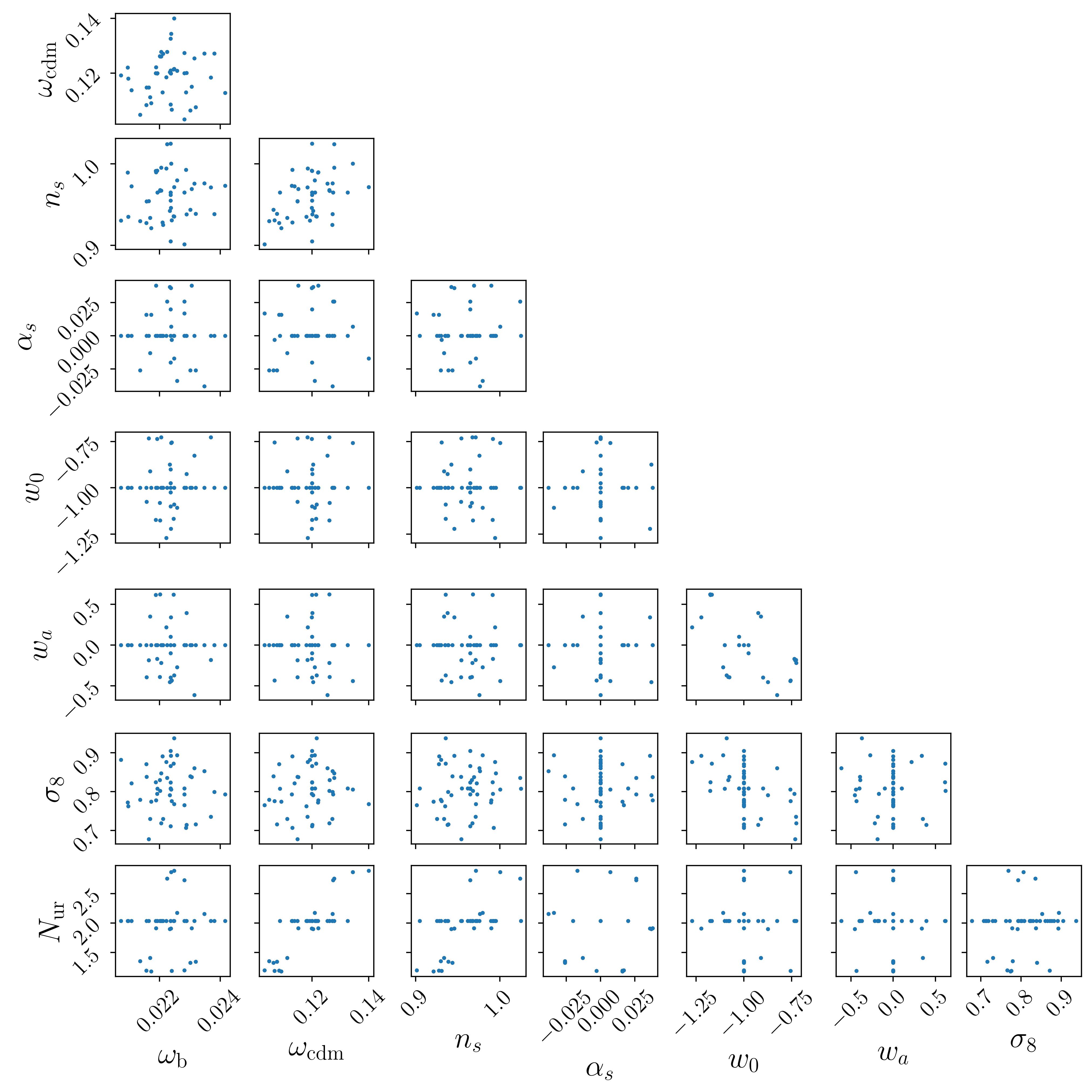}
\caption{Parameter distribution of all available \texttt{AbacusSummit} cosmologies.}
\label{fig:abacus_paramters}
\end{figure}

\section{AbacusSummit light cone simulations}
\label{sec:AbacusSummit}

To build a model which predicts our summary statistics, we create an emulator, which is calibrated on \abacussummit \citep{Maksimova2021}, 
a suite of cosmological simulations that were run with the \texttt{Abacus} N-body code \citep{Garrison2019, Garrison2021}. The \textit{base} \abacussummit\ simulations follow the evolution of $6912^3$ dark matter particles in a periodic box of $2\Gpch$ on a side, with a mass resolution of $2.1 \times 10^9 \Msunh$. Group finding is performed using the \texttt{CompaSO} hybrid Friends-of-Friends/Spherical Overdensity algorithm \citep{Hadzhiyska2022}.

Beyond the periodic simulations, \cite{Hadzhiyska2022} presented catalogues of dark matter halos in lightcones that are constructed from the base simulation boxes, which can be used to build realistic mock catalogues on the sky that also capture the evolution of clustering across redshift. The lightcones cover an octant of the sky up to a redshift of 0.8 and span 63 different cosmologies within an extended $w_0w_a$CDM parameter space (see Fig.~\ref{fig:abacus_paramters}),
\begin{equation}
\label{eq:abacus_parameters}
    \Theta_{\rm \texttt{AbacusSummit}} = \{ \omegac, \omegab, \sigma_8, n_\mathrm{s}, \nrun, \Neff, w_0, w_a \} \,,
\end{equation} 
where $\omegac$ and $\omegab$ are the physical cold dark matter and baryon densities, $\sigma_8$ is the amplitude of density fluctuations, $n_\mathrm{s}$ is the spectral index of the primordial power spectrum, $\nrun$ is the running of the spectral index, $\Neff$ is the effective number of ultra-relativistic species, $w_0$ is the present-day dark energy equation of state, and $w_a$ captures the evolution of dark energy across redshift. The simulations assume zero curvature, and the value of the Hubble parameter $h$ was calibrated to match the Cosmic Microwave Background angular size of the acoustic scale, $\theta_*$, as observed by \citetalias{Planck2020}.

The 63 cosmologies span from $5 \sigma$ to $8 \sigma$ around the best-fit cosmology from Planck 2018 \citep[][hereafter \citetalias{Planck2020}]{Planck2020}, assuming base-$\Lambda$CDM. This fiducial cosmology is referred to in this work as \texttt{c000}.  For the fiducial cosmology \texttt{c000}, 25 phase realizations with different initial conditions are available, which we later use to correct for cosmic variance \cite{Racz2023}. Only one phase realization is available for all the other cosmologies, referred to as \texttt{ph00}.

On top of the halo lightcones, shear information ($\kappa$ and $\gamma$) is provided in terms of shear maps with resolution $A_\mathrm{pix}=0.046\,\mathrm{arcmin}^2$ $(\texttt{nside}=16384)$ at ascending redshift slices from $0.1 < z < 2.45$, with $\Delta z = 0.05$ steps. We use these shear maps to build a reduced shear $g=\gamma/(1-\kappa)$ catalogue that matches the UNIONS-like $p(z)$. This is done by randomly selecting $w_i N_\mathrm{pix}$ pixels from the corresponding shear map, where the weights $w_i$ result from integrals over the UNIONS-like $p(z)$ shown in Fig.~\ref{fig:nz} up to redshift of one and $N_\mathrm{pix}$ are the available pixels at redshift one. We are cutting the UNIONS-like $p(z)$ at $z = 1$ because, due to the lightcones' geometry, the sky's initial octant reduces above a redshift of 0.8. We decided to lose some signal by neglecting the high redshift galaxies for higher emulator accuracy for the lensing statistics.

\begin{table*}
    \renewcommand{\arraystretch}{1.2}
    \centering
    \caption{List of cosmological and HOD parameters used in our analysis. For each, we quote the parameter symbol, the prior distribution, and the physical interpretation. We note that the prior distribution for all parameters is uniform, except the baryon density, for which we adopt a normal distribution with a mean and dispersion as specified.}
    \begin{tabular}{| l | l | c | l|}
        \hline
        Parameter & Prior distribution  & Interpretation\\
        \hline
        $\omega_{\rm b}$ & $\unif [0.021, 0.024]$  &  Physical baryon density\\
        $\omega_{\rm cdm}$ & $\unif [0.103, 0.140]$  & Physical cold dark matter density\\
        $\sigma_8$ & $\unif [0.678, 0.937]$ & Amplitude of matter fluctuations in $8\,h^{-1}{\rm Mpc}$ spheres\\
        $n_\mathrm{s}$ & $\unif [0.901, 1.025]$  & Spectral index of the primordial power spectrum\\
        $\nrun$ & $\unif [-0.038, 0.038]$ & Running of the spectral index\\
        $\Neff$ & $\unif [2.190, 3.902]$  & Number of ultra-relativistic species\\
        $w_0$ & $\unif [-1.271, -0.726]$  & Present-day dark energy equation of state\\
        $w_a$ & $\unif [-0.615, 0.621]$  & Time evolution of the dark energy equation of state\\
        \hline
        $\log_{10} M_{\rm cut}$ & $\unif [12, 13.5]$  & Minimum halo mass to host a central in units of $M_\odot/h$\\
        $\log_{10} M_1$ & $ \unif[13, 14]$  & Typical halo mass to host one satellite in units of $M_\odot/h$\\
        $\log_{10} \sigma$ & $\unif [-5.0, 0.0]$ & Slope of the transition from hosting zero to one central\\
        $\alpha$ & $\unif [0.0, 3.0]$  & Power-law index for the mass dependence of the number of satellites \\
        $\kappa$ & $\unif [0.0, 3.0]$  & Parameter that modulates the minimum halo mass to host a satellite\\
        $B_{\rm cen}$ & $\unif [-1.0, 1.0]$  & Environment-based assembly bias for centrals\\
        $B_{\rm sat}$ & $\unif [-2.0, 2.0]$  & Environment-based assembly bias for satellites \\
        $s$ & $\unif [-2.0, 2.0]$  & Modulation of the satellite profile \\
        \hline
    \end{tabular}
    \label{tab:priors}
\end{table*}

\subsection*{Galaxy-halo connection model}
\label{subsec:hod}

We model the connection between dark matter haloes and galaxies using an extended HOD prescription. This model statistically assigns galaxies to haloes based on their properties, most importantly, the host halo mass.

The first ingredient of the HOD model \citep{Zheng2007} is the expectation of finding a galaxy at the center of a halo of mass $M$,
\begin{align}
    \langle N_{\rm c} \rangle(M) = \frac{1}{2} \left[1 + \mathrm{erf} \left(\frac{\mathrm{log}_{10} M - \mathrm{log}_{10} M_{\rm cut}}{\sqrt{2} \sigma} \right)  \right]\,
    \label{eq:Ncen}
\end{align}
where $\mathrm{erf}(x)$ is the error function, $M_{\rm cut}$ is the minimum mass required to host a central, and $\sigma$ controls the slope of the transition between having zero to one central. Additionally to the central galaxy, the expected number of satellite galaxies is given by
\begin{align}
    \langle N_{\rm s} \rangle(M) = \langle N_{\rm c} \rangle(M) \left(\frac{M - \kappa M_{\rm cut}}{M_1} \right)^{\alpha}\,
    \label{eq:Nsat}
\end{align}
where $\kappa M_{\rm cut}$ is the minimum mass required to host a satellite, $M_1$ is the typical mass to host one satellite, and $\alpha$ is the power law index for the number of galaxies.

Motivated by \cite{Yuan2022}, we allow an extension to the base model to account for environment-based secondary bias\footnote{In the literature, this parameter is also called assembly bias.}, captured by the parameters $B_{\rm cen}$ and $B_{\rm sat}$, which modulate the effective number of centrals and satellites, respectively, to have a dependence on the local environment,
\begin{align}
    \mathrm{log}_{10} M_\mathrm{cut}^\mathrm{eff} &= \mathrm{log}_{10} M_\mathrm{cut} + B_\mathrm{cen}(\delta^\mathrm{rank} - 0.5)\, , \\
    \mathrm{log}_{10} M_\mathrm{1}^\mathrm{eff} &= \mathrm{log}_{10} M_\mathrm{1} + B_\mathrm{sat}(\delta^\mathrm{rank} - 0.5) \, ,
\end{align}
where the environment is defined as the smoothed matter density around the halo centres, using a top-hat filter of radius $R_s = 3\,h^{-1}{\rm Mpc}$. These environments are then assigned to `ranks' $\delta^\mathrm{rank} \in [0,1]$, with environments denser than the median having $\delta^\mathrm{rank}>0.5$. The base model with no secondary bias is recovered when $B_{\rm cen} = B_{\rm sat} = 0$.

In our HOD model, the positions of the satellites are assigned to randomly selected halo particles by default. We relax this assumption by assigning a probability $p_i$ to each halo particle to host a satellite galaxy \citep{Yuan2018},
\begin{equation}
    p_i = \frac{\langle N_{\rm s} \rangle(M) }{N_\mathrm{p}}\left[1+s\left(1-\frac{2\,R_i}{N_\mathrm{p}-1}\right)\right]\, ,
\end{equation}
where $N_\mathrm{p}$ is the number of particles in the halo, $R_i$ is the rank of the particle ordered by distance to the halo centre such that the outermost particle has rank 0 and the innermost particle has ranking $N_\mathrm{r}-1$. $s$ is a free parameter to modulate this behaviour. In particular, an $s>0$ results in satellite distribution being puffier than the halo profile and $s<0$ to satellite distribution being more concentrated. We summarize all cosmological and HOD parameters in Table \ref{tab:priors}. We note that we restrict the stated HOD parameters in Table \ref{tab:priors} to those parameters where the HOD model produces a number density that is between 0.9 and five times the number density of LOWZ between $0.15<z<0.4$. In Fig.~\ref{fig:abacus_paramters_HOD}, we see that this means that all points with $\log_{10} M_\mathrm{cut}\lesssim 12.8$ are removed. We further decided not to downsample the HOD catalogues to match exactly the LOWZ $p(z)$ and number of galaxies and instead down-sample such that the shape of the LOWZ $p(z)$ is achieved. This means that HODs that create more galaxies have a higher number density then the real data. In our analysis, we do not take into account if the exact number density is achieved, but we created an emulator that learns the non-uniform prior range, such that we ensure we run the analysis only in the trained range. 

We populate the halo lightcones with galaxies using \abacushod\footnote{\url{https://abacusutils.readthedocs.io/en/latest/hod.html}} \citep{Yuan2022}, a highly efficient HOD sampling code that contains a wide range of HOD variations. When generating the HOD catalogues, we ignore the potential redshift evolution of the HOD parameters. Implementing a redshift-dependent HOD modelling is outside the scope of this work, but it is an aspect we plan to explore in future work in light of upcoming Stage-IV cosmological experiments.

\section{Estimating a covariance matrix}
\label{sec:T17}
To perform a parameters forecast analysis and compare our emulator accuracy to the measurement noise, we need to estimate a covariance matrix that describes the noise in the LOWZxUNIONS data.
To estimate this covariance matrix, we use the simulations described in \cite[][hereafter \citetalias{Takahashi2017}]{Takahashi2017}, because \abacussummit\ does not provide enough realizations to measure a LOWZxUNIONS-like covariance matrix. 
The \citetalias{Takahashi2017} simulations trace the nonlinear evolution of $2048^3$ particles across a series of nested cosmological volumes. These volumes initially have a side length of $L=450 \Mpch$ at low redshift, progressively increasing in size at higher redshifts, resulting in 108 distinct full-sky realizations. These simulations were conducted using the Gadget-3 $N$-body code \citep{Springel2005} and are publicly accessible\footnote{\citetalias{Takahashi2017} simulations: \url{http://cosmo.phys.hirosaki-u.ac.jp/takahasi/allsky_raytracing/}}. The cosmological parameters governing matter and vacuum energy density are $\Omega_{\rm m}=1-\Omega_\Lambda=0.279$, the baryon density parameter $\Omega_{\rm b}=0.046$, the dimensionless Hubble constant $h=0.7$, the power spectrum normalization $\sigma_8=0.82$, and the spectral index $n_{\rm s}=0.97$. Information on shear in the \citetalias{Takahashi2017} simulations is provided through 108 full-sky realizations of $\gamma$ and $\kappa$, each divided into 38 ascending redshift slices, each having the same comoving distance. 
Since the correlation functions are measured down to $1'$, we made use of those realizations that have a pixel resolution of $A_\mathrm{pix}=0.18\,\mathrm{arcmin}^2$ ($\texttt{nside}=8192$). 

To replicate the preliminary UNIONS-like $p(z)$ depicted in Fig.~\ref{fig:nz}, we constructed a weighted average from the first 30 $\gamma$ and $\kappa$ redshift slices. The weight for each redshift slice is determined by integrating the $p(z)$ across the corresponding width of the redshift slice. From these stacked shear maps, we created galaxy shear catalogues based by extracting the shear information at the galaxy positions of the preliminary UNIONS catalogue (v1.0 ShapePipe), covering roughly a footprint of $3500\,\mathrm{deg}^2$. To add shape noise, we combined the two-component reduced shear $g = \gamma/(1-\kappa)$ of each object with a shape noise contribution, $\epsilon^{\mathrm{s}}$, to create observed ellipticities ${\boldsymbol\epsilon}^{\mathrm{obs}}$ \citep{Seitz1997}, as follows:
\begin{equation}
\epsilon^{\mathrm{obs}} = \frac{{ \epsilon}^{\mathrm{s}}+{ g}}{1+{ \epsilon}^\mathrm{s}{g}^*} \, ,
\label{eq:ebos}
\end{equation}
where asterisk `$*$' indicates complex conjugation. The noise contribution, $\epsilon^{\mathrm{s}}$, is derived from the measured ellipticities of the preliminary UNIONS catalogue by randomly rotating the observed ellipticities to erase the underlying correlated shear signal. This ensures that the distribution of $|\epsilon^{\mathrm{s}}|$ matches the distribution of the measured ellipticities. Furthermore, accounting for the corresponding weight $w$ for each shape measurement ensures we use the correct effective number density.

To model the foreground galaxies, we use a standard HOD description (Eq.~\ref{eq:Ncen} and Eq.~\ref{eq:Nsat}) without secondary bias or satellite profile modulation. To determine the parameters, we started from $\mathrm{log}_{10} M_\mathrm{cut}=13.29$, $\sigma=0.55$, $\mathrm{log}_{10} M_1=14.18$, $\kappa = 0.14$ and $\alpha=1.55$, which are taken from \cite{Harnois-Deraps:2018} and agree with the results from \cite{Kobayashi2022}. From there, we adjusted the parameters such that the angular clustering measured from the real LOWZ data and T17 were matched. This resulted in the end to $\mathrm{log}_{10} M_\mathrm{cut}=13.20$, $\sigma=0.55$, $\mathrm{log}_{10} M_1=14.26$, $\kappa = 0.14$ and $\alpha=1.55$. In contrast with the model described in Sect.~\ref{subsec:hod}, we down-sampled it so that the $p(z)$ and the total number of galaxies match those from the LOWZ data (Fig.~\ref{fig:nz}) to ensure that the shot noise is accurately modelled in the covariance matrix. 
Given that UNIONS measures only in the northern hemisphere, we also restrict the LOWZ data to declinations larger than $24.8^\circ$, resulting in an effective area of $3530\,\mathrm{deg}^2$.  

\begin{figure}
\includegraphics[width=\linewidth]{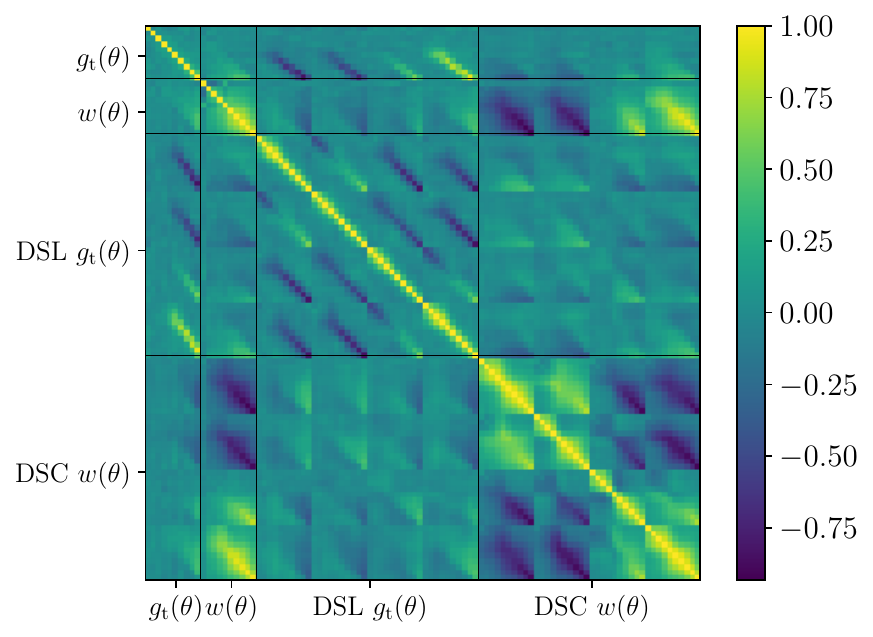}
\caption{Correlation matrix measured from the 756 \citetalias{Takahashi2017} realizations for all four summary statistics. It is clearly seen that the lower two quantiles are anti-correlated with the upper two quantiles and that the lensing statistics are uncorrelated with clustering statistics.}
\label{fig:correlation_matrix}
\end{figure}

Lastly, to increase the number of mock data, we rotate the galaxy positions and ellipticities seven times by $50^\circ$ along the lines of constant declination. This procedure created 756 almost independent mock catalogues from which the covariance is measured. The resulting correlation matrix is shown in Fig.~\ref{fig:correlation_matrix}. The next section gives a detailed description of the data vectors' measurements.

\section{Measuring the summary statistics}
\label{sec:model_measurement}
We aim to build an emulator that is able to predict second-order and higher-order statistics, which are calibrated on \abacussummit\. For the second-order statistics, we measure the galaxy-galaxy lensing signal $g_\mathrm{t}(\theta)$ and the angular clustering signal $w(\theta)$, together with their corresponding DSS measurements. To measure the correlation functions we use \texttt{treecorr} \citep{Jarvis:2004} and measure the second-order statistics in ten logarithmic bins between $1'< \theta < 200'$. While $g_\mathrm{t}(\theta)$ is the standard correlation function between foreground galaxies and the shear of background galaxies, the angular clustering $w(\theta)$ is estimated using \cite{Landy1993},
\begin{equation} \label{eq:CCF}
    w(\theta) = \frac{\rm \mathcal{GG} - 2\mathcal{GR} + \mathcal{RR}}{\rm \mathcal{RR}} \,,
\end{equation}
where $\mathcal{GG}$, $\mathcal{GR}$ and $\mathcal{RR}$ are the normalized galaxy-galaxy, galaxy-random, and random-random pair counts.

For the measurement of the DSS, we follow Sec.~\ref{Aperture_stat}, and smooth the projected galaxy density fields with a Gaussian of $10'$. We also correct the resulting aperture number as described in Eq.~\eqref{eq:Nap_corrected}, which is redundant for the \abacussummit\ but important for the \citetalias{Takahashi2017} measurements as masks are incorporated in them. We then divide the aperture number field into five quantiles, where only those pixels with $99\%$ coverage are used\footnote{For \citetalias{Takahashi2017} measurements the threshold is $80\%$ given that these have masks.}. Using again \texttt{treecorr}, these quantile points are either correlated with the shear of the background galaxies, resulting in five density-split lensing (DSL) signals, or with the foreground galaxy positions, resulting in five density-split clustering (DSC) signals. The shear signals of DSL are measured in ten logarithmic bins between $5'< \theta < 300'$, and for the clustering signals of the DSC, we use again ten logarithmic bins between $1'< \theta < 200'$. For DSC, we measure the angular cross-correlation function between the quintile positions and the galaxy field \cite{Landy1993},
\begin{equation} \label{eq:CCF}
    w^\mathrm{DS}(\theta) = \frac{\mathcal{QG} - \mathcal{GR} - \mathcal{QR} + \mathcal{RR}}{ \mathcal{RR}} \,,
\end{equation}
where $\mathcal{QG}$ and $\mathcal{QR}$ are the normalized quantile-galaxy and quantile-random pair counts. Since the five quantiles are not linearly independent, we need to drop one of the quantiles to avoid an inevitable covariance matrix. Since the middle quantile has the lowest signal-to-noise, we decided to ignore that quantile.  

We set the number of random query points or unique pixels inside the survey area to be ten times the number of foreground galaxies, and we used all source galaxies for our calculations without adding shape noise. For the \citetalias{Takahashi2017} measurements, we use all source galaxies and 100 times the number of foreground galaxies as query points. To decrease the contribution from shot noise to our model vectors, we created HOD realizations using $n_\mathrm{seeds} = 10$ different random seeds for the same HOD parameters\footnote{Since the HOD is a probabilistic model, the choice of random seed can impact the resulting number of galaxies in a halo, as well as the assigned positions of satellite galaxies within that halo.}, and we averaged our statistics across these different HOD realizations,
\begin{equation}
    \bar{\boldsymbol{m}}^\texttt{ph00}(\Theta) = \frac{1}{n_\mathrm{seeds}} \sum_{s=1}^{n_\mathrm{seeds}} \boldsymbol{m}_s^\texttt{ph00}(\Theta) \, ,
    \label{eq:model1}
\end{equation}
where \texttt{ph00} stands for the phase realization available for all cosmologies. 
Next, to decrease the bias arising from having only one phase realization for all cosmologies except \texttt{c000}, we compute a correction vector based on the $n_\mathrm{p}=25$ phase realizations provided for \texttt{c000}. 
\begin{equation}
    \boldsymbol{m}^\mathrm{cor}(\Theta;\texttt{h}) =   \frac{n_\mathrm{seeds}}{n_\mathrm{p}}\frac{\sum_{p=1}^{n_\mathrm{p}}  \boldsymbol{m}^p(\texttt{c000};\texttt{h})}{\sum_{s=1}^{n_\mathrm{seeds}} \boldsymbol{m}_s^\texttt{ph00}(\texttt{c000};\texttt{h})} \,  \, ,
    \label{eq:model2}
\end{equation}
where the upper index runs over all the 25 phase realizations, the lower index over the ten different HOD random seed realizations, and \texttt{h} stands for the reference HOD parameters. 
The corrected model results to
\begin{equation}
    \bar{\boldsymbol{m}}(\Theta,\texttt{h}) = \boldsymbol{m}^\mathrm{cor}(\Theta;\texttt{h}) \, \bar{\boldsymbol{m}}^\texttt{ph00}(\Theta)
    \label{eq:model3}
\end{equation}
Since the corrected model depends strongly on the used reference HOD parameters \texttt{h}, we built an average over the $n_\mathrm{\texttt{h}}=10$ HOD parameters chosen at random from the parameter training nodes, which results in the final model
\begin{equation}
    \boldsymbol{m}(\Theta) = \frac{1}{n_\mathrm{\texttt{h}}} \sum_{\texttt{h}=1}^{n_\mathrm{\texttt{h}}} \bar{\boldsymbol{m}}(\Theta,\texttt{h}) \, .
    \label{eq:model4}
\end{equation}

\begin{figure}
\centering
\includegraphics[width=\columnwidth]{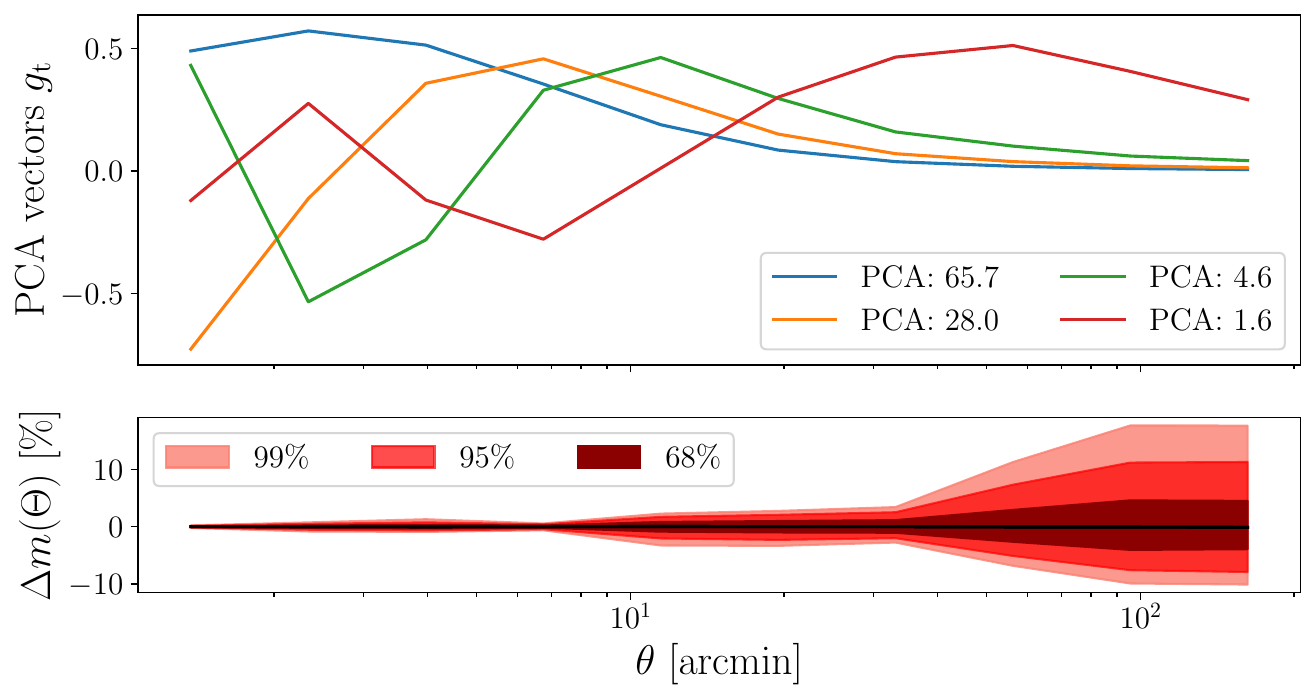}
\caption{The upper panel shows the first four significant eigenvectors and the corresponding eigenvalues in the legend. The shape of eigenvector one closely matches the shape of $g_\mathrm{t}$. The lower panel shows a relative difference between the PCA cleaned and raw model vectors. The DSL, DSC and $w(\theta)$ behaviour are very similar.}
\label{fig:PCA_correction}
\end{figure}

As a final step, we clean the measured models using a principal component analysis (PCA) as a final noise reduction \cite{EE2021,Arico2021}. The idea is that those eigenvectors that correspond to larger eigenvalues describe the largest deviation in the model due to the variation of cosmological or HOD parameters. The eigenvectors corresponding to the smaller eigenvalues mostly describe the remaining noise in the data. Therefore, by transforming the model $\boldsymbol{m}(\Theta)$ to the PCA space and discarding all eigenvectors with eigenvalues $\lesssim 5$ and then transforming back, we reduce some of the noise in the model vectors as seen in Fig.~\ref{fig:PCA_correction}. We tested the case where we did not apply the PCA compression and found that this slightly increases the precision in our parameter constraints. However, as we cannot ensure if this apparent increase in precision comes from smoothing out true or noisy features, we decided to proceed with the conservative approach of performing the PCA cleaning. 

\section{Building the emulator}
\label{sec:emulator}

This section describes the procedure from the cleaned measurements to the emulator. We also test the emulator's accuracy and discuss possible improvements. In particular, we test whether more HOD training nodes are helpful, whether the assumed reference HOD for the correction is essential, and whether shot noise is the dominant noise source.

As described above, we have 63 cosmologies and 100 HOD per cosmology to train the emulator. In the following plots, we discard the 100 measurements from the training set that belong to \texttt{c000} and use them solely for testing the emulator. Using the remaining 6200 measurements, we built individual emulators for $g_\mathrm{t}$, $w(\theta)$ and the DSS analogies. We built on the neural network emulator CosmoPower \citep{COSMOPOWER2022} for the emulation, using four layers of 256 nodes each to avoid fitting the remaining noise in the models. We have tested many other settings for the number of layers and nodes but found no improvement. Furthermore, the software \texttt{optuna}\footnote{\url{https://optuna.org/}} gave only minor improvements while being more time-consuming to optimize for every emulator. We therefore continued with the chosen setup.

\begin{figure}
\includegraphics[width=\columnwidth]{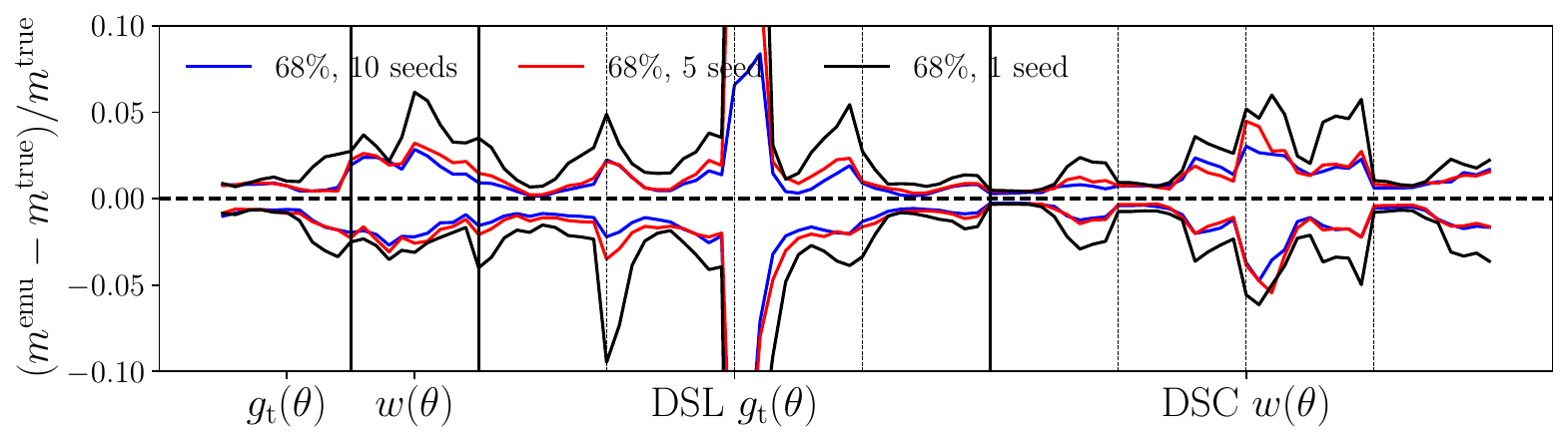}
\caption{We show the emulator accuracy using different numbers of random seeds to build the average in Eq.~\eqref{eq:model1} and Eq.~\eqref{eq:model2}.}
\label{fig:seeds_ratio}
\end{figure}

\begin{figure}
\includegraphics[width=\columnwidth]{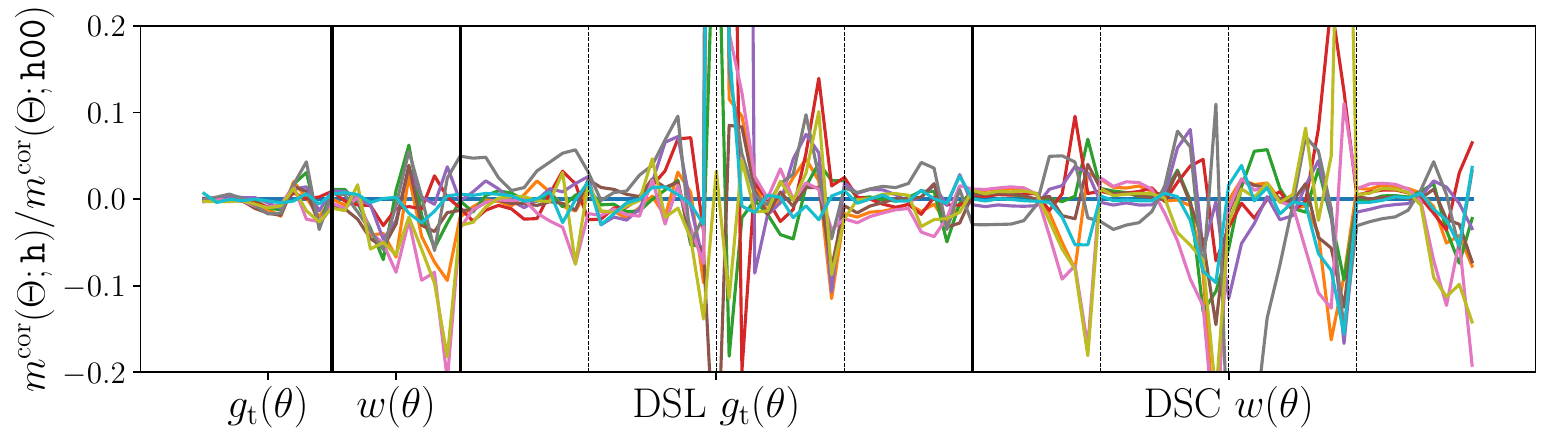}
\caption{We show the ratio between the correction factors built with different realizations of HOD parameters with respect to \texttt{h00}.}
\label{fig:hod_correction}
\end{figure}

\begin{figure}
\includegraphics[width=\columnwidth]{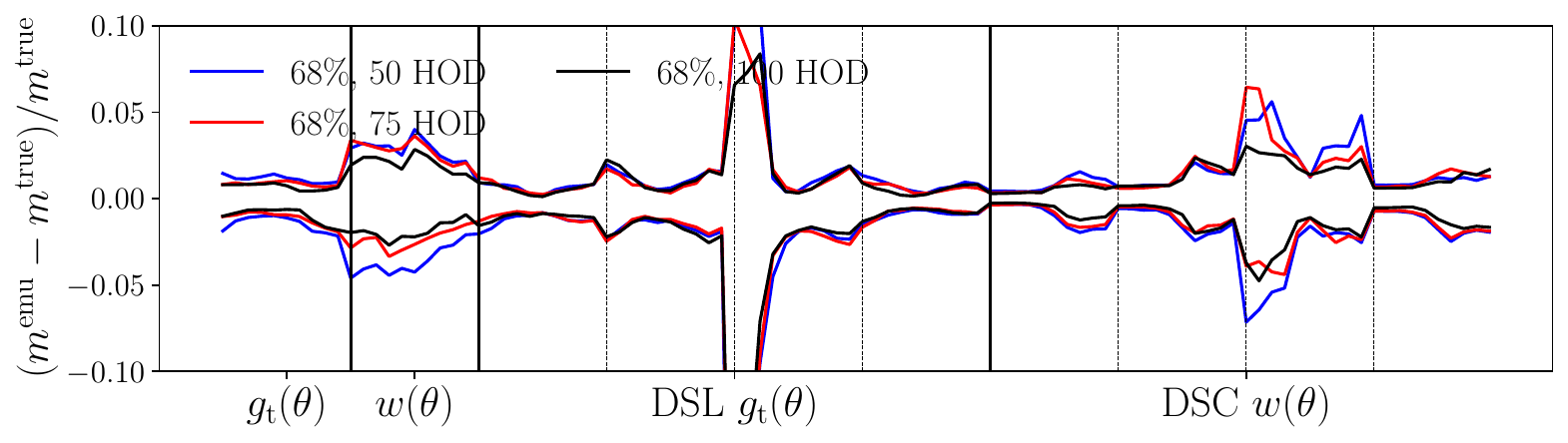}
\caption{Emulator accuracy if trained on less than 100 HOD parameters per cosmology.}
\label{fig:hod_acc}
\end{figure}

Our first finding is that a large contribution to the noise for the current setup comes from the different random seeds for the HOD of the foreground galaxies. As discussed above, we used ten noise seeds to populate the lens galaxies in the HOD process. It is seen in Fig.~\ref{fig:seeds_ratio} that using only one seed is not enough and that with five noise seeds, the accuracy starts to converge. It seems that more noise seeds might help decrease the remaining noise further. Next, we test if the assumed HOD parameters for the cosmic variance correction in Eq.~\eqref{eq:model4} are essential. In Fig.~\ref{fig:hod_correction}, we compare the correction vectors relative to the $\boldsymbol{m}^\mathrm{cor}(\Theta;\texttt{h00})$ correction vector. Although the resulting accuracy of the emulator does not change if only one set of models is used or multiple sets are averaged first, it does affect the overall bias in the model by $\sim 10\%$ for elements of the model vectors that are close to zero. We, therefore, continue using the average of all ten model sets. Lastly, we tested whether more HOD training nodes are helpful. For this, we reduced the number of HOD training nodes from 100 to 75 and 50. As shown in Fig.~\ref{fig:hod_acc}, the accuracy of the clustering statistics significantly improves by using more HOD parameters. Although the difference between 100 and 75 HOD per cosmology is insignificant, there is still some improvement in using slightly more HOD per cosmology.

\begin{figure}
\includegraphics[width=\columnwidth]{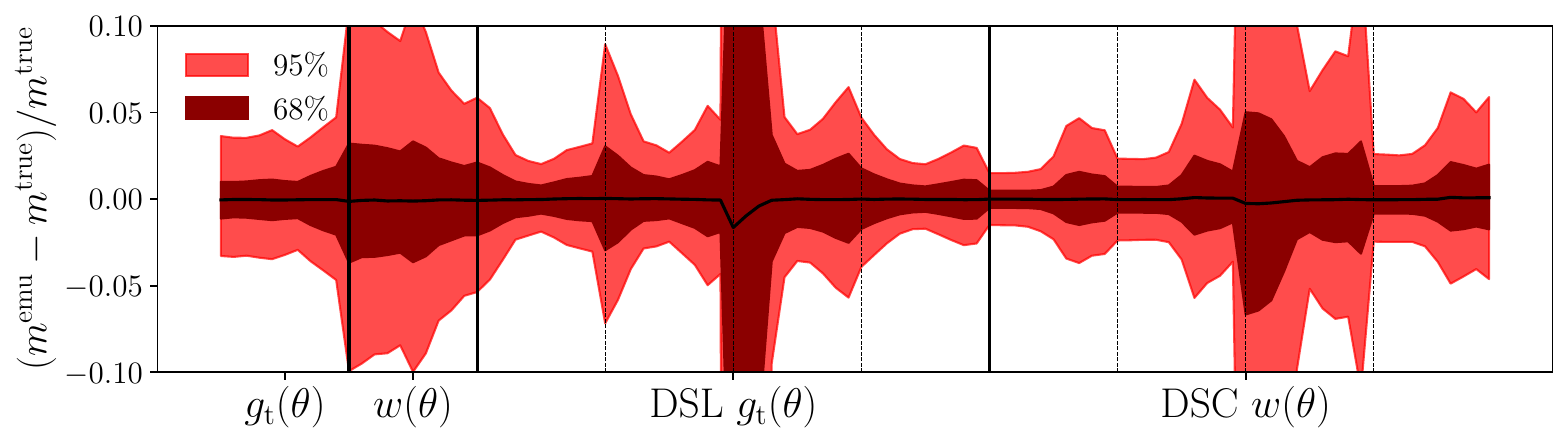}
\includegraphics[width=\columnwidth]{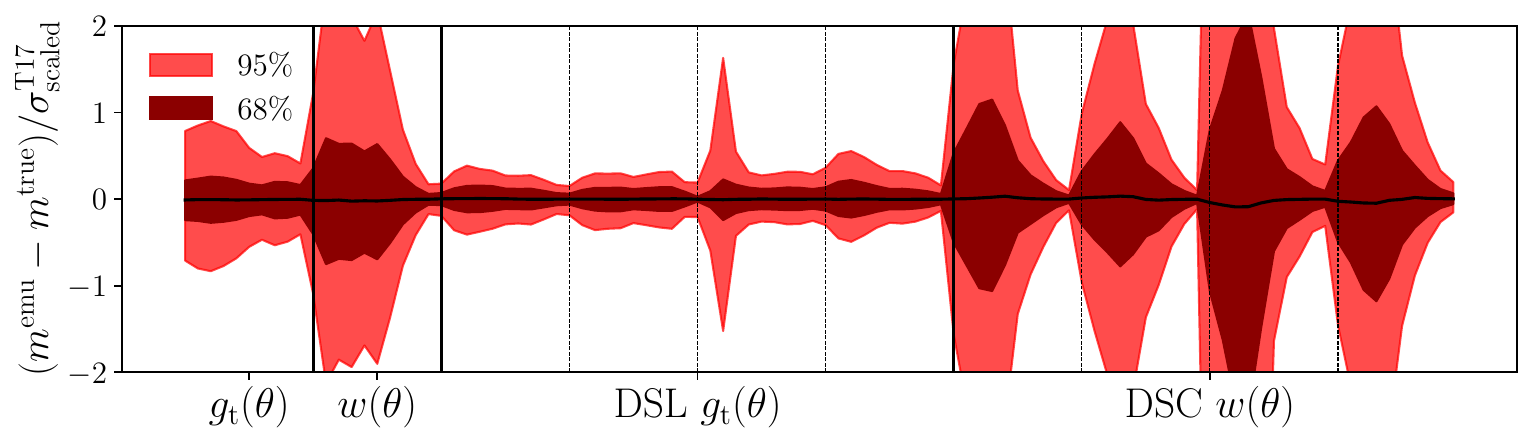}
\caption{Accuracy of the emulator using leave-one-out method. The upper panel shows the relative accuracy, and the middle panel shows the accuracy relative to the expected LOWZxUNIONS-like uncertainty described in Sect.~\ref{sec:T17}. The lower panel is the accuracy relative to the cosmic variance noise together with the emulator noise defined in Eq.\eqref{eq:emu_cov}. The black dashed line shows the mean deviation.}
\label{fig:emu_acc}
\end{figure}

\begin{figure}
\includegraphics[width=\columnwidth]{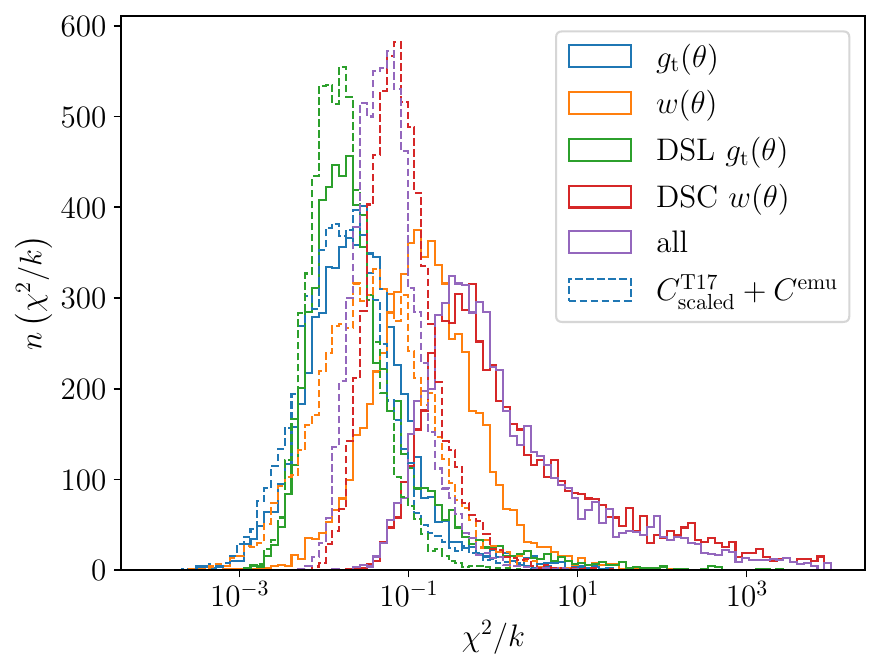}
\caption{Emulator accuracy if trained on less than 100 HOD parameters per cosmology.}
\label{fig:emu_acc_chi2}
\end{figure}

Finally, to quantify the overall error in the model, we perform the leave-one-out method. In particular, we successively leave out one of the 63 cosmologies and train on the other 62. We show the overall accuracy in Fig.~\ref{fig:emu_acc}, where the upper panels show the residuals and the lower panel the difference divided by the square root of the diagonal of the scaled \citetalias{Takahashi2017} covariance matrix. To create the scaled covariance matrix, we mean that take the ratio of 
\begin{equation}
    r_i(\Theta) = \frac{\boldsymbol{m}_i(\Theta)}{1/n_\mathrm{res} \sum_j^{n_\mathrm{res}} d_i^j} \, ,
\end{equation}
where $n_\mathrm{res}=756$ is the number of \citetalias{Takahashi2017} data vectors $d^{j}$ that is used to compute the covariance matrix and build the outer product with the $C^\mathrm{T17}$ covariance 
\begin{equation}
    C_{ij,\mathrm{scaled}}^\mathrm{T17}(\Theta) = C_{ij}^\mathrm{T17} r_i(\Theta) r_j(\Theta)
    \label{eq:scaling_cov}
\end{equation}
We notice that, especially for the DSC, the emulator's error is in the same order as the expected  LOWZxUNIONS-like noise. To account for inaccuracies in the emulator, we estimate a covariance matrix $C^\mathrm{emu}$ that describes the scatter in quantity $\Delta m = m^\mathrm{emu}-m^\mathrm{true}$. We then add this covariance matrix to the covariance matrix that describes the expected noise in the data 
\begin{equation}
    \label{eq:emu_cov}
    C = C^\mathrm{T17}_\mathrm{scaled} + C^\mathrm{emu} \,.
\end{equation}
Since the residual plots are misleading in quantifying the accuracy of the emulator alone due to cross-correlations of the angular bins, we measure
\begin{equation}
    \chi^2/k = \frac{1}{k} \Delta m^\mathrm{T} C^{-1} \Delta m,
\end{equation}
where $k$ is the degrees-of-freedom, which is the total number of elements since we are not fitting $m^\mathrm{emu}$ to $m^\mathrm{true}$. The distribution of $\chi^2/k$ for the combined and individual statistics are shown in Fig.~\ref{fig:emu_acc_chi2}. The two lensing statistics peak well below one, but as expected, the clustering statistics have a significant number of $\chi^2/k>1$. However, if we add the emulator accuracy $C^\mathrm{emu}$, the clustering statistics also move towards $\chi^2/k\ll 1$. The emulator is more accurate for points closer to the center of the parameter space, and the biggest deviations are for large $\log_{10} M_\mathrm{cut}$.

\section{Forecast}
\label{sec:forecast}
With the emulator and covariance matrix described in the previous chapter, we focus now on a parameter constraint forecast. Since the estimated covariance matrix $C$ is a random variable, we follow \cite{Percival2021} to compute the likelihood. Given a data vector $\boldsymbol{d}$ and covariance matrix $C$ of rank $n_\mathrm{d}$ measured from $n_\mathrm{r}$ realizations, the posterior distribution of a model vector $\boldsymbol{m}(\Theta)$ that depends on $n_{\Theta}$ parameters is
\begin{equation}
    \boldsymbol{P}\left(\boldsymbol{m}(\Theta)|\boldsymbol{d},C\right) \propto |C|^{-\frac{1}{2}} \left( 1 + \frac{\chi^2}{n_{\rm r}-1}\right)^{-m/2}\, ,
    \label{eq:t_distribution}
\end{equation}
where
\begin{equation}
\chi^2 =  \left[\boldsymbol{m}(\Theta)-\boldsymbol{d}\right]^{\rm T} C^{-1} \left[\boldsymbol{m}(\Theta)-\boldsymbol{d}\right] \, .
\label{eq:chi2}
\end{equation}
The power-law index $m$ is 
\begin{equation}
    m = n_\Theta+2+\frac{n_\mathrm{r}-1+B(n_\mathrm{d}-n_\Theta)}{1+B(n_\mathrm{d}-n_\Theta)}
\end{equation}
with $n_{\rm d}$ being the number of data points and
\begin{equation}
    B = \frac{n_\mathrm{r}-n_\mathrm{d}-2}{(n_\mathrm{r}-n_\mathrm{d}-1)(n_\mathrm{r}-n_\mathrm{d}-4)} \, .
    \label{eq:B}
\end{equation}
By setting $m=n_\mathrm{r}$, the formalism of \cite{Sellentin2016} is recovered. To estimate the likelihood in Eq.~\eqref{eq:t_distribution}, we make use of a Markov-Chain Monte Carlo process\footnote{\url{https://github.com/justinalsing/affine}}. As the reference data vector $\boldsymbol{d}$, we use the emulator to predict a vector at cosmological and HOD parameters in Table \ref{tab:fiducial}. We also tested a broader set of parameters and found similar results independent of the chosen reference parameters, as long as these are sufficiently inside the prior range. We also scaled the \citetalias{Takahashi2017} covariance matrix using the procedure explained in Eq.~\eqref{eq:scaling_cov}. We decided to vary all HOD and all cosmological parameters except the running of the spectral index, the time evolution of the dark energy equation of state, and the number of ultra-relativistic species. Furthermore, we use Gaussian priors for $\omegab$ and $n_\mathrm{s}$ around the truth and standard deviation given by \citetalias{Planck2020}. We exclude all elements of the analysis that do not follow a Gaussian distribution, which we have checked using the Shapiro-Wilk test based on the \citetalias{Takahashi2017} mock data \citep{SHAPIRO}. Although we varied in $\omegac$, $\omegab$ and $\sigma_8$, we used the fact that $\Omegam = (\omegac+\omegab+\omega_\mathrm{ncdm})/h^2$, where $\omega_\mathrm{ncdm}=6.442 \times 10^{-4}$, and that $S_8 = \sigma_8 \sqrt{\Omegam/0.3}$. The dimensionless Hubble parameter $h$ is determined by using another emulator, which is constructed using the fact the acoustic scale matches to \citetalias{Planck2020} (see Sect.\,\ref{sec:AbacusSummit}).

\begin{figure*}[ht]
\includegraphics[width=\linewidth]{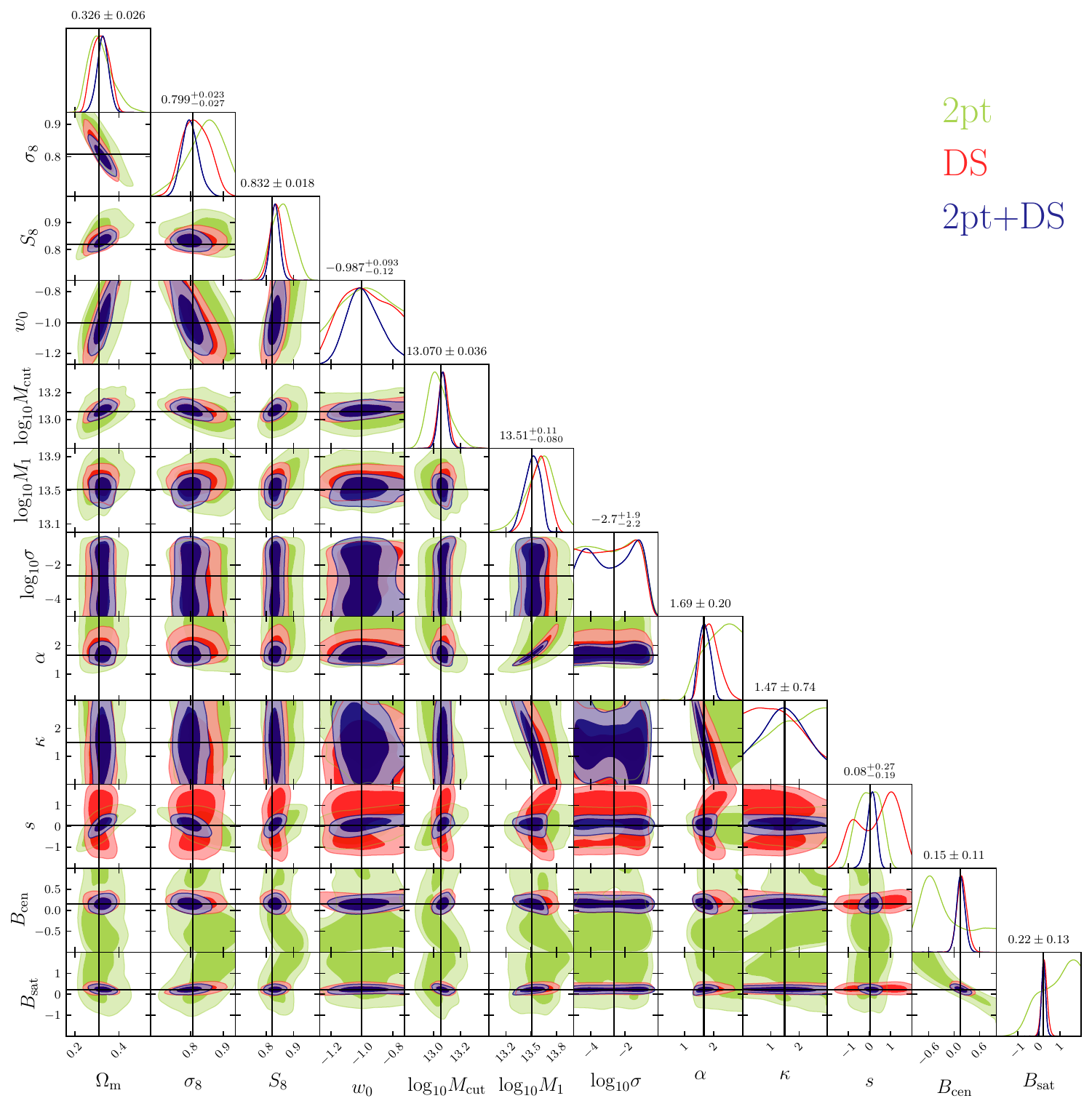}
\caption{Parameter constraints comparing second-order statistics to the DSS. The stated parameter constraints belong to the combined analysis of weak lensing and galaxy clustering.}
\label{fig:MCMC_2pt_vs_DS}
\end{figure*}

\begin{figure*}[ht]
\includegraphics[width=\linewidth]{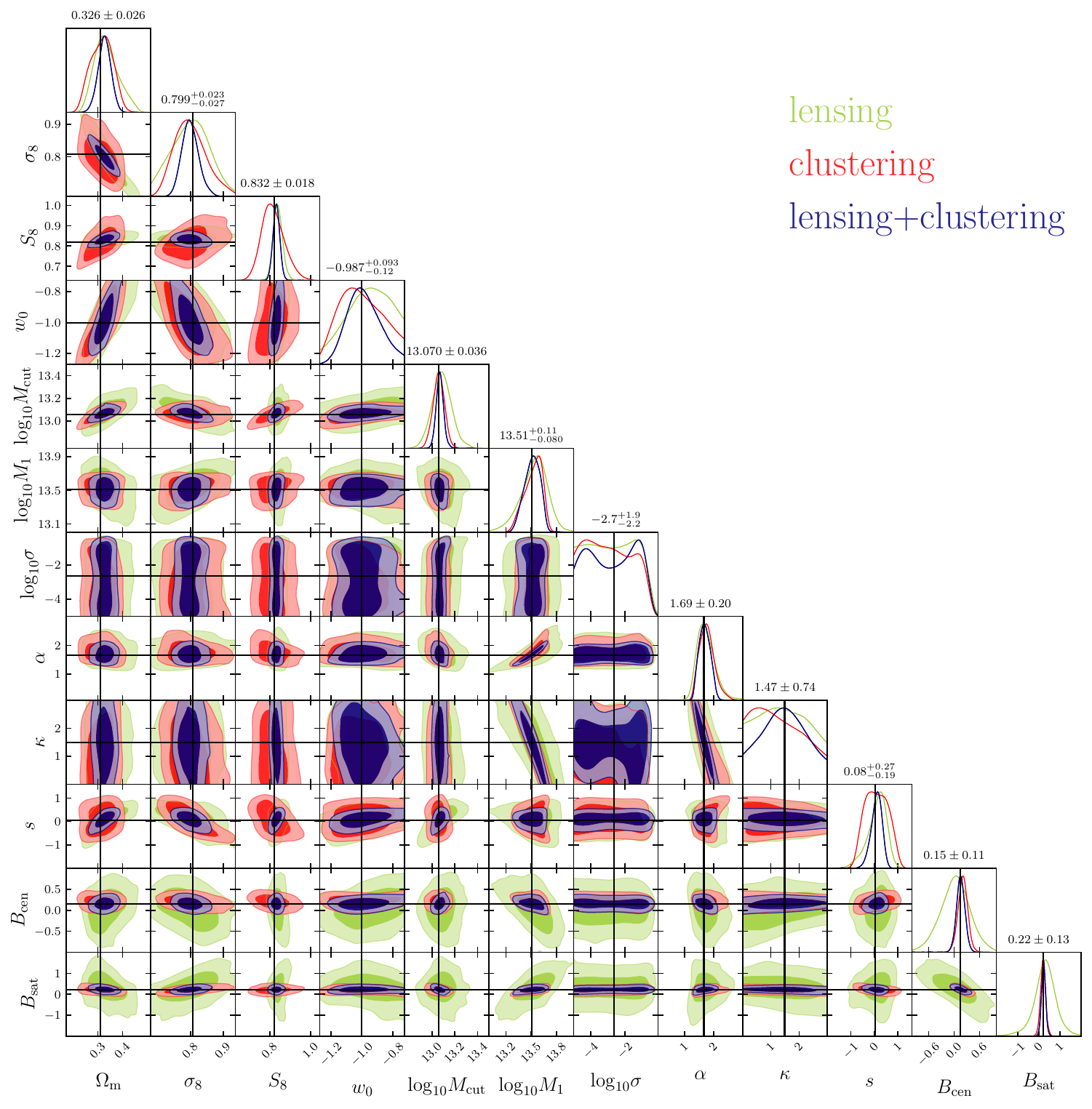}
\caption{Parameter constraints comparing clustering to lensing statistics. The stated parameter constraints belong to the combined analysis of two-point and the DSS.}
\label{fig:MCMC_lensing_vs_clustering}
\end{figure*}

The first results are shown in Fig.~\ref{fig:MCMC_2pt_vs_DS}, where we compare the constraints using only second-order statistics to DSS. It is seen that DSS are more constraining for all cosmological and HOD parameters, and adding second-order statistics does not help much in gaining constraining power. Second-order statistics are only more constraining for the $s$-parameter that modulates the satellite profile, affecting the small scales that are smeared out for the DSS. However, combining DSS with second-order DSS improves the constraining power for almost all parameters since combining different statistics helps break degeneracies in the multi-dimensional posterior distribution. The much higher constraining power on the assembly bias parameters using the DSS is particularly interesting. We expect this comes from the fact that the assembly bias itself is an environment-dependent parameter. The comparison between clustering and lensing in Fig.~\ref{fig:MCMC_lensing_vs_clustering} shows that DSC is responsible for the strong constraining power in the assembly bias parameters. Lensing alone significantly outperforms clustering only for the $S_8$ parameter. This partly comes from the fact we are measuring our statistics in the projected sky and ignoring redshift-space distortions. Combining clustering and lensing improves the constraining power for all the remaining parameters. We have tested that the constraining power stays approximately constant independent of where we are probing the posterior in parameter space as long none of the parameters is bounded by the prior range. Lastly, we validated that we get unbiased results using a non-PCA cleaned prediction as a reference data vector. 

\section{Conclusion}
\label{sec:conclusion}
This paper presents a simulation-based modelling of density split statistics, combining clustering (DSC) and lensing (DSL). Furthermore, we combine these DSS with second-order statistics of angular clustering $w(\theta)$ and galaxy-galaxy lensing $g_\mathrm{t}(\theta)$. The latter correlates the shear of background galaxies with the position of foreground galaxies. We build our model based on the \texttt{AbacusSummit} simulations, where we sampled the foreground galaxies using a halo occupation distribution (HOD). Using these simulations, we build an emulator and test several measurement choices that can help improve the emulator's accuracy. Our first finding is that the emulator accuracy for the shear measurements ($g_\mathrm{t}(\theta)$ and DSL) is sufficient compared to the expected noise when combining the position of the LOWZ galaxies with the shear from preliminary UNIONS data. However, for the clustering statistics, the emulator accuracy is sometimes of the same order as the expected noise of the LOWZ data. Since the relative accuracy is similar for both clustering and lensing statistics, this poorer performance comes from the significantly smaller noise of the clustering statistics. To account for the noise in the emulator, we derive an emulator covariance matrix using the leave-one-out method. For the parameter forecast analysis, we add this emulator noise to the LOWZxUNIONS-like covariance matrix measured from \cite{Takahashi2017} simulations. With the model and covariance, we then perform a parameter forecast.

Our first finding is that compared to second-order statistics, DSS is more constraining for cosmological and HOD parameters. Second-order statistics are beneficial to constrain satellite modulation parameters, as this information is smeared out for the DSS. When comparing clustering and lensing statistics, we find that, as expected, lensing is more constraining in $S_8$, and DSC is responsible for the constraining power of the assembly bias parameter. We note that the HOD parameter $\kappa$ is bound by the priors. For a similar analysis in the future, we have to widen the prior range for this parameter to avoid informative priors.
Furthermore, we have ignored the effects of intrinsic alignment (IA) between lens and source galaxies, which can be problematic if sources and lenses overlap significantly in redshift. Since IA is not easy to incorporate into our model, we recommend using only source and lens galaxies that are well separated in redshift. Another advantage is that magnification bias and boost correction can be ignored if one considers only well-separated source and lens redshift distributions. Lastly, we also note that the stated parameter constraints might increase if shear bias and uncertainty in the source redshift distribution are included in the analysis, which we ignored for this analysis. 

Overall, we conclude that the \texttt{AbacusSummit} simulations are a powerful tool to derive higher-order statistic models that can constrain cosmological and HOD parameters. The \texttt{AbacusSummit} simulations provide an excellent opportunity to validate results from analytical models in the literature, derive better constraining powers, and shed light on new aspects of the galaxy halo connection. We look forward to applying the emulator to data from weak lensing surveys such as UNIONS and Euclid.

\begin{acknowledgements}

We thank the anonymous referee for their fruitful comments. We want to thank Mike Jarvis for developing and maintaining \texttt{treecorr}, and Alessio Mancini for developing \texttt{CosmoPower} emulator, which provided a significant part of our model. Furthermore, we thank the authors of \texttt{AbacusSummit} and \texttt{AbacusHOD} for providing and maintaining these excellent simulations and code.

We thank the UNIONS team for providing preliminary shear catalogues and redshift distributions, which allowed us to perform a realistic parameter forecast.

We acknowledge financial support from the Canadian Space Agency (Grants 21EXPCOI3 and 23EXPROSS1), the Waterloo Centre for Astrophysics and the National Science and Engineering Research Council Discovery Grants program.

This research was enabled in part by support provided by Calcul Quebec\footnote{\url{https://docs.alliancecan.ca/wiki/Narval}} and Compute Ontario\footnote{\url{https://docs.alliancecan.ca/wiki/Graham}} and the Digital Research Alliance of Canada\footnote{\url{alliancecan.ca}}. This research used resources from the National Energy Research Scientific Computing Center, supported by the Office of Science of the U.S. Department of Energy under Contract No. DE-AC02-05CH11231.
\end{acknowledgements}
\appendix

\section{Additional material}
\begin{figure}
\includegraphics[width=\linewidth]{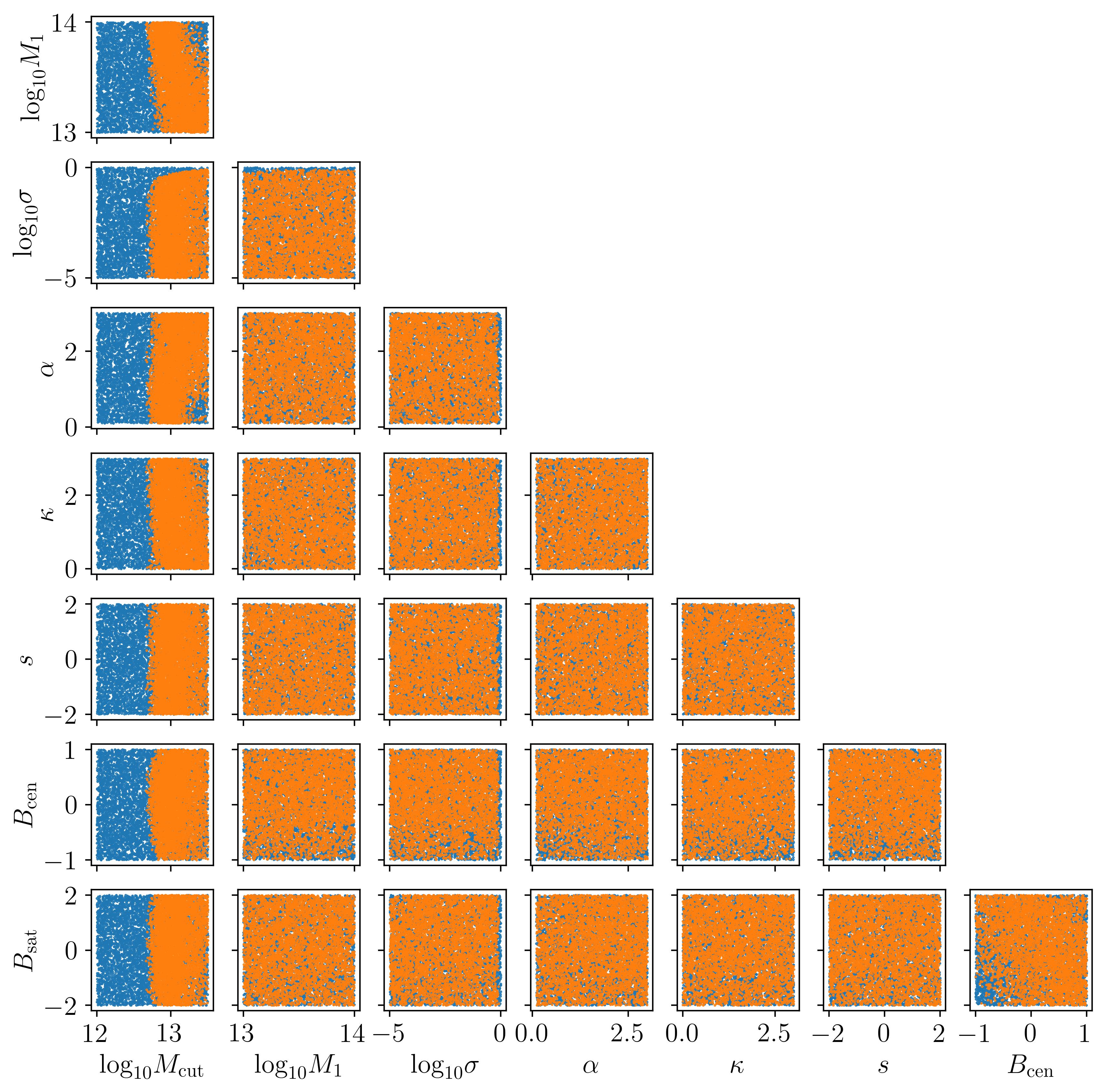}
\caption{Parameter distribution of the initial considered HOD parameters in blue. In orange, we show those points used to build the emulator, as they have a number density between 0.9 and five times the density of LOWZ.}
\label{fig:abacus_paramters_HOD}
\end{figure}

\begin{figure*}[ht]
\includegraphics[width=\linewidth]{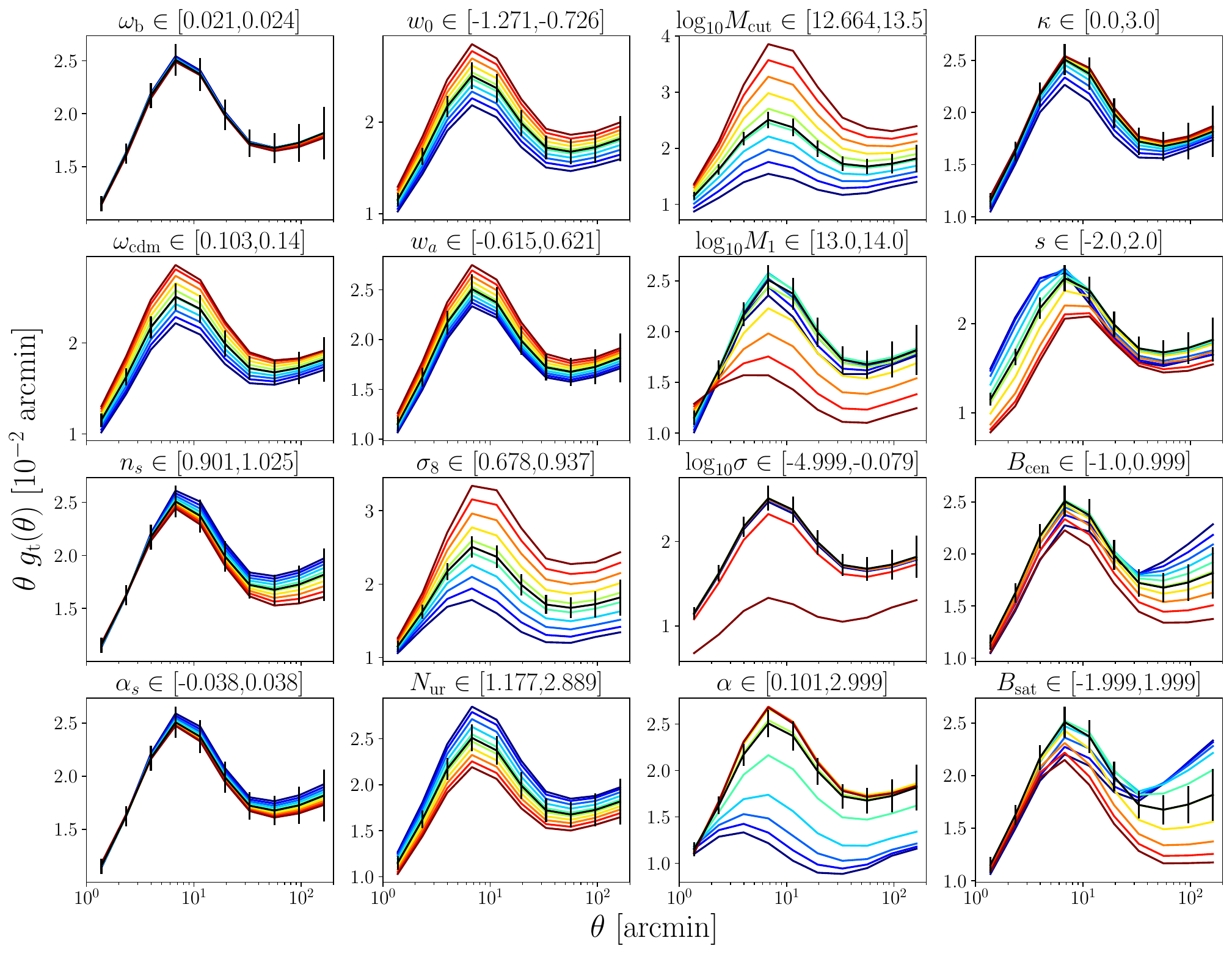}
\caption{Model response of $g_\mathrm{t}(\theta)$ on cosmological and HOD parameters. In each panel, only the stated parameter in the title is varied, and all the other parameters are fixed to \texttt{c000} and \texttt{h00}. The black dashed line with error bars is the model at parameters stated in Table \ref{tab:fiducial}, where the error bars come from the scaled \citetalias{Takahashi2017} covariance matrix.}
\label{fig:gammat_response}
\end{figure*}

\begin{figure*}[ht]
\includegraphics[width=\linewidth]{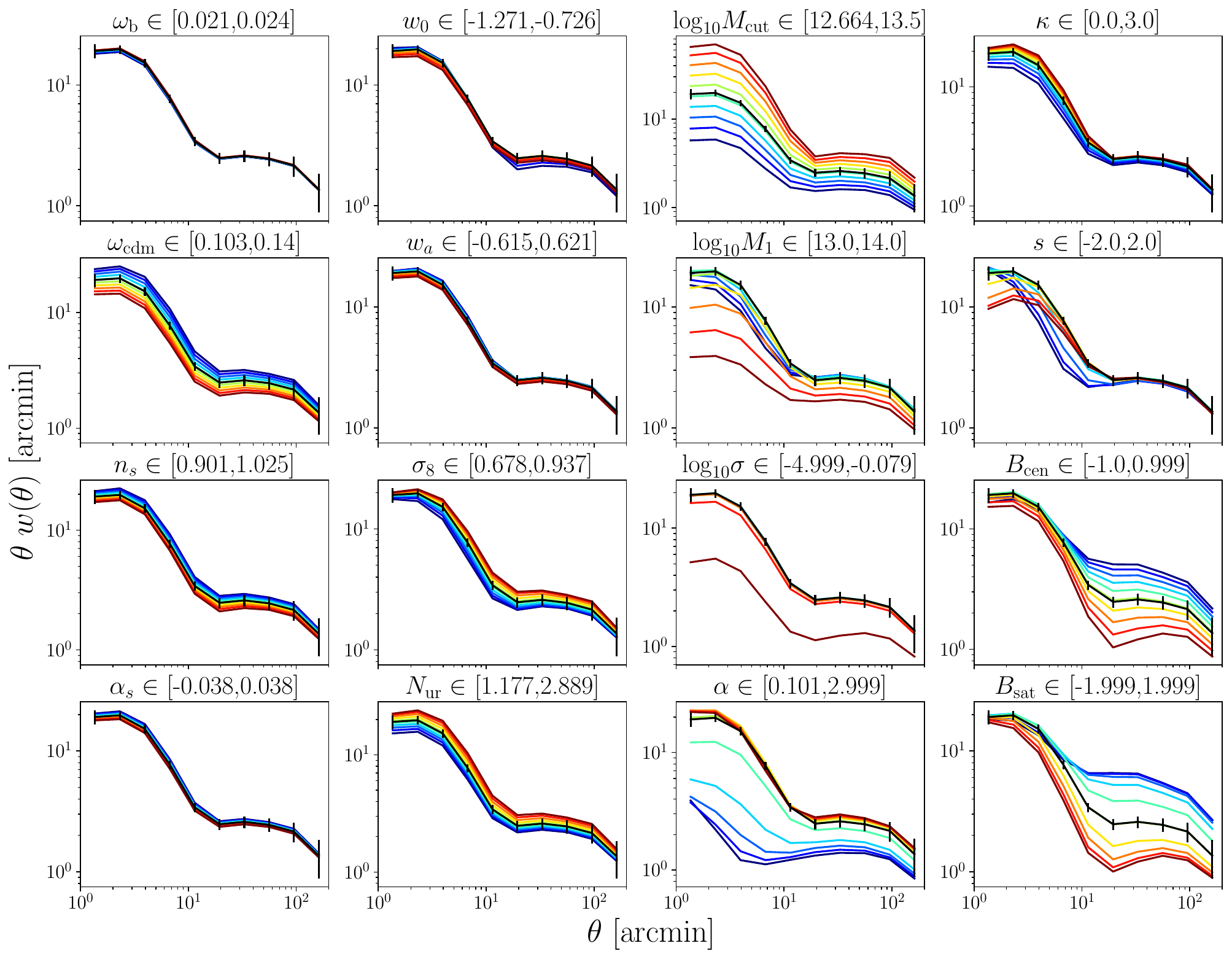}
\caption{Same as Fig.~\ref{fig:gammat_response} but for $w(\theta)$.}
\label{fig:wtheta_response}
\end{figure*}

\begin{figure*}[ht]
\includegraphics[width=\linewidth]{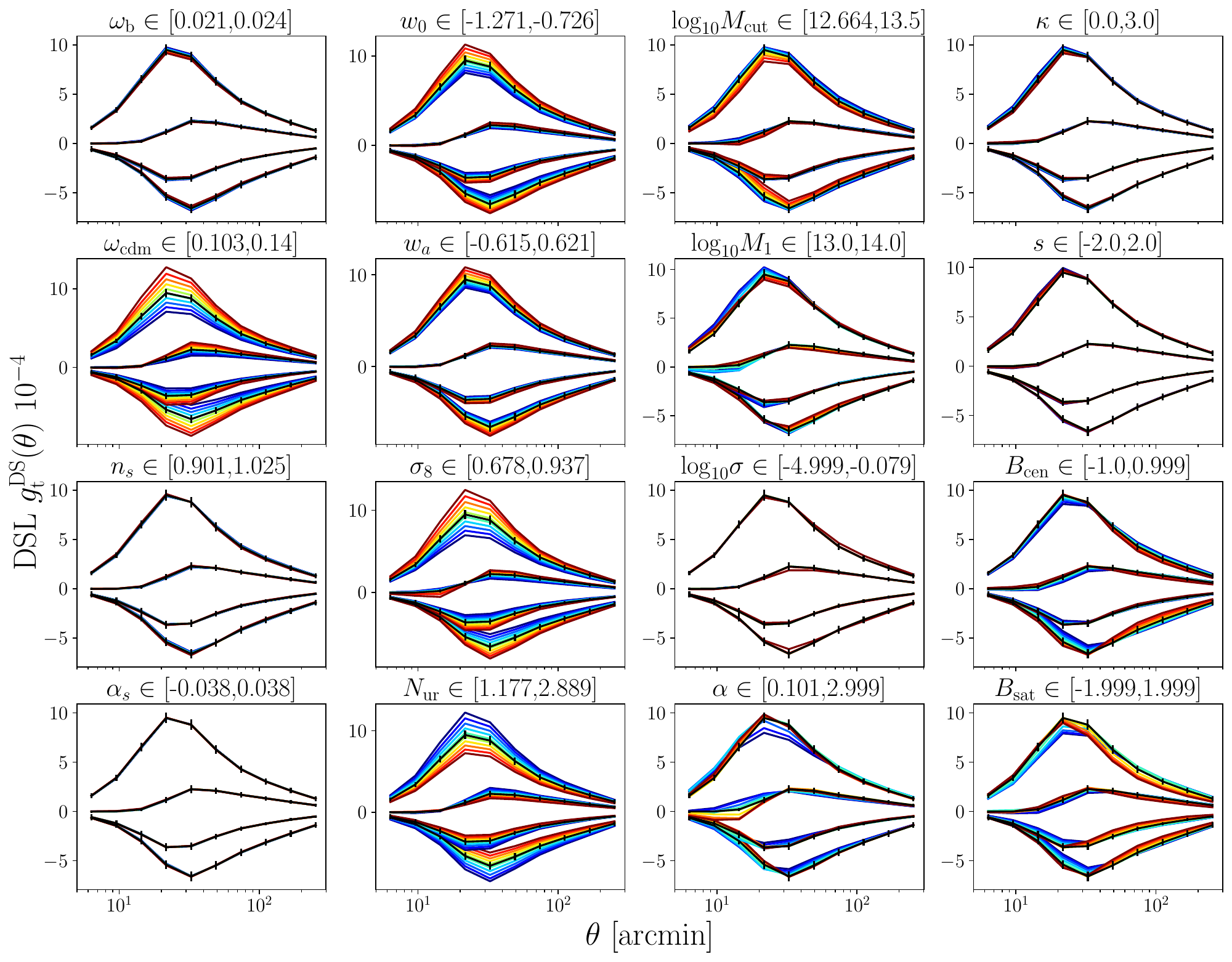}
\caption{Same as Fig.~\ref{fig:gammat_response} but for DSL.}
\label{fig:DSL_response}
\end{figure*}

\begin{figure*}[ht]
\includegraphics[width=\linewidth]{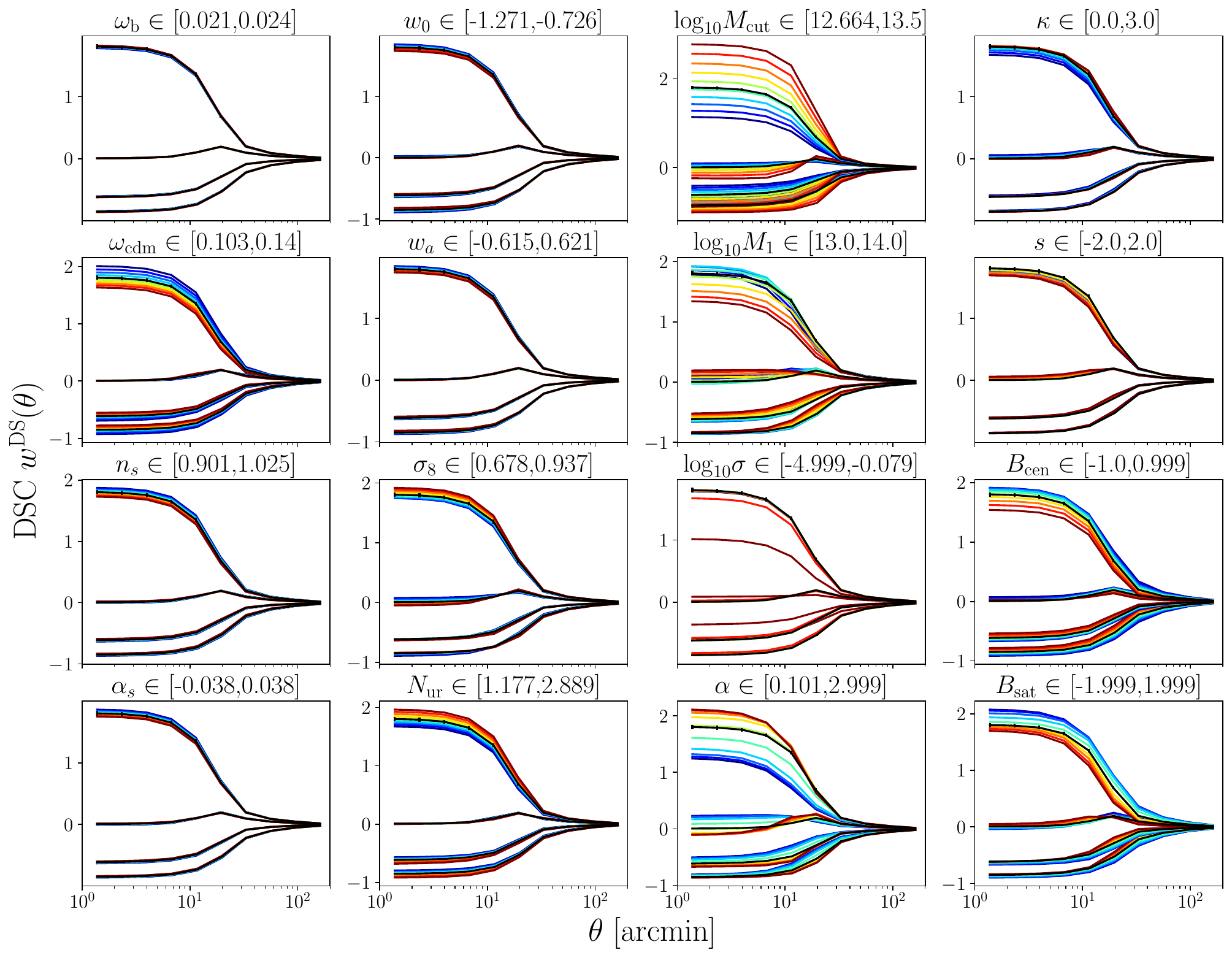}
\caption{Same as Fig.~\ref{fig:gammat_response} but for DSC.}
\label{fig:DSC_response}
\end{figure*}

In this chapter, we collect plots that are complementary to this work. In Fig.~\ref{fig:abacus_paramters_HOD}, we show the parameter distribution of the HOD parameters. In blue, we show the initial guess of parameters, and in orange, those that have not created too many or too few galaxies compared to the LOWZ data. In Fig.~\ref{fig:gammat_response} to Fig.~\ref{fig:DSC_response}, we show the response of our model if only one parameter is changed, while the others are fixed to the fiducial cosmology and HOD given in Table \ref{tab:fiducial}. We also show the model for the fiducial cosmology and HOD in every plot in black, together with the expected LOWZxUNIONS-like noise.

\begin{table*}
    \renewcommand{\arraystretch}{1.2}
    \centering
    \caption{Fiducial cosmological and HOD parameters.}
    \begin{tabular}{| l | l | l | l | l | l | l | l | l | l | l | l | l | l | l | l | l |}
        \hline
        Parameter & $\omega_{\rm b}$ & $\omega_{\rm cdm}$ & $\sigma_8$ & $n_\mathrm{s}$ & $\nrun$ & $\Neff$ & $w_0$ & $w_a$ \\
        \hline
        values & $0.0224$  & $0.12$ & $0.8076$  & $0.9649$  & $0.0$  & $3.046$  & $-1.0$  & $0.0$    \\
        \hline
        \hline
        Parameter & $\mathrm{log}_{10} M_{\rm cut}$ & $\mathrm{log}_{10} M_1$ & $\mathrm{log}_{10} \sigma$ & $\alpha$ & $\kappa$ & $B_{\rm cen}$ & $B_{\rm sat}$ & $s$ \\
        \hline
        values   & $13.0566$   & $13.5086$  & $-2.6466$  & $1.6686$  & $1.4868$   & $0.0415$  & $0.1461$ & $0.2189$   \\
        \hline
    \end{tabular}
    \label{tab:fiducial}
\end{table*}

\bibliographystyle{apsrev4-1}
\bibliography{bibliography}

\begin{thebibliography}{55}%
\makeatletter
\providecommand \@ifxundefined [1]{%
 \@ifx{#1\undefined}
}%
\providecommand \@ifnum [1]{%
 \ifnum #1\expandafter \@firstoftwo
 \else \expandafter \@secondoftwo
 \fi
}%
\providecommand \@ifx [1]{%
 \ifx #1\expandafter \@firstoftwo
 \else \expandafter \@secondoftwo
 \fi
}%
\providecommand \natexlab [1]{#1}%
\providecommand \enquote  [1]{``#1''}%
\providecommand \bibnamefont  [1]{#1}%
\providecommand \bibfnamefont [1]{#1}%
\providecommand \citenamefont [1]{#1}%
\providecommand \href@noop [0]{\@secondoftwo}%
\providecommand \href [0]{\begingroup \@sanitize@url \@href}%
\providecommand \@href[1]{\@@startlink{#1}\@@href}%
\providecommand \@@href[1]{\endgroup#1\@@endlink}%
\providecommand \@sanitize@url [0]{\catcode `\\12\catcode `\$12\catcode
  `\&12\catcode `\#12\catcode `\^12\catcode `\_12\catcode `\%12\relax}%
\providecommand \@@startlink[1]{}%
\providecommand \@@endlink[0]{}%
\providecommand \url  [0]{\begingroup\@sanitize@url \@url }%
\providecommand \@url [1]{\endgroup\@href {#1}{\urlprefix }}%
\providecommand \urlprefix  [0]{URL }%
\providecommand \Eprint [0]{\href }%
\providecommand \doibase [0]{http://dx.doi.org/}%
\providecommand \selectlanguage [0]{\@gobble}%
\providecommand \bibinfo  [0]{\@secondoftwo}%
\providecommand \bibfield  [0]{\@secondoftwo}%
\providecommand \translation [1]{[#1]}%
\providecommand \BibitemOpen [0]{}%
\providecommand \bibitemStop [0]{}%
\providecommand \bibitemNoStop [0]{.\EOS\space}%
\providecommand \EOS [0]{\spacefactor3000\relax}%
\providecommand \BibitemShut  [1]{\csname bibitem#1\endcsname}%
\let\auto@bib@innerbib\@empty
\bibitem [{\citenamefont {{Euclid Collaboration}}\ \emph
  {et~al.}(2023)\citenamefont {{Euclid Collaboration}}, \citenamefont
  {{Ajani}}, \citenamefont {{Baldi}}, \citenamefont {{Barthelemy}},
  \citenamefont {{Boyle}}, \citenamefont {{Burger}}, \citenamefont {{Cardone}},
  \citenamefont {{Cheng}}, \citenamefont {{Codis}}, \citenamefont {{Giocoli}}
  \emph {et~al.}}]{HOWLS2023}%
  \BibitemOpen
  \bibfield  {author} {\bibinfo {author} {\bibnamefont {{Euclid
  Collaboration}}}, \bibinfo {author} {\bibfnamefont {V.}~\bibnamefont
  {{Ajani}}}, \bibinfo {author} {\bibfnamefont {M.}~\bibnamefont {{Baldi}}},
  \bibinfo {author} {\bibfnamefont {A.}~\bibnamefont {{Barthelemy}}}, \bibinfo
  {author} {\bibfnamefont {A.}~\bibnamefont {{Boyle}}}, \bibinfo {author}
  {\bibfnamefont {P.}~\bibnamefont {{Burger}}}, \bibinfo {author}
  {\bibfnamefont {V.~F.}\ \bibnamefont {{Cardone}}}, \bibinfo {author}
  {\bibfnamefont {S.}~\bibnamefont {{Cheng}}}, \bibinfo {author} {\bibfnamefont
  {S.}~\bibnamefont {{Codis}}}, \bibinfo {author} {\bibfnamefont
  {C.}~\bibnamefont {{Giocoli}}},  \emph {et~al.},\ }\href {\doibase
  10.1051/0004-6361/202346017} {\bibfield  {journal} {\bibinfo  {journal}
  {\aap}\ }\textbf {\bibinfo {volume} {675}},\ \bibinfo {eid} {A120} (\bibinfo
  {year} {2023})},\ \Eprint {http://arxiv.org/abs/2301.12890}
  {arXiv:2301.12890} \BibitemShut {NoStop}%
\bibitem [{\citenamefont {{Martinet}}\ \emph {et~al.}(2018)\citenamefont
  {{Martinet}}, \citenamefont {{Schneider}}, \citenamefont {{Hildebrandt}},
  \citenamefont {{Shan}}, \citenamefont {{Asgari}}, \citenamefont {{Dietrich}},
  \citenamefont {{Harnois-D{\'e}raps}}, \citenamefont {{Erben}}, \citenamefont
  {{Grado}}, \citenamefont {{Heymans}} \emph {et~al.}}]{Martinet:2018}%
  \BibitemOpen
  \bibfield  {author} {\bibinfo {author} {\bibfnamefont {N.}~\bibnamefont
  {{Martinet}}}, \bibinfo {author} {\bibfnamefont {P.}~\bibnamefont
  {{Schneider}}}, \bibinfo {author} {\bibfnamefont {H.}~\bibnamefont
  {{Hildebrandt}}}, \bibinfo {author} {\bibfnamefont {H.}~\bibnamefont
  {{Shan}}}, \bibinfo {author} {\bibfnamefont {M.}~\bibnamefont {{Asgari}}},
  \bibinfo {author} {\bibfnamefont {J.~P.}\ \bibnamefont {{Dietrich}}},
  \bibinfo {author} {\bibfnamefont {J.}~\bibnamefont {{Harnois-D{\'e}raps}}},
  \bibinfo {author} {\bibfnamefont {T.}~\bibnamefont {{Erben}}}, \bibinfo
  {author} {\bibfnamefont {A.}~\bibnamefont {{Grado}}}, \bibinfo {author}
  {\bibfnamefont {C.}~\bibnamefont {{Heymans}}},  \emph {et~al.},\ }\href
  {\doibase 10.1093/mnras/stx2793} {\bibfield  {journal} {\bibinfo  {journal}
  {\mnras}\ }\textbf {\bibinfo {volume} {474}},\ \bibinfo {pages} {712}
  (\bibinfo {year} {2018})},\ \Eprint {http://arxiv.org/abs/1709.07678}
  {arXiv:1709.07678} \BibitemShut {NoStop}%
\bibitem [{\citenamefont {{Harnois-D{\'e}raps}}\ \emph
  {et~al.}(2021)\citenamefont {{Harnois-D{\'e}raps}}, \citenamefont
  {{Martinet}}, \citenamefont {{Castro}}, \citenamefont {{Dolag}},
  \citenamefont {{Giblin}}, \citenamefont {{Heymans}}, \citenamefont
  {{Hildebrandt}},\ and\ \citenamefont {{Xia}}}]{Harnois-Deraps:2021}%
  \BibitemOpen
  \bibfield  {author} {\bibinfo {author} {\bibfnamefont {J.}~\bibnamefont
  {{Harnois-D{\'e}raps}}}, \bibinfo {author} {\bibfnamefont {N.}~\bibnamefont
  {{Martinet}}}, \bibinfo {author} {\bibfnamefont {T.}~\bibnamefont
  {{Castro}}}, \bibinfo {author} {\bibfnamefont {K.}~\bibnamefont {{Dolag}}},
  \bibinfo {author} {\bibfnamefont {B.}~\bibnamefont {{Giblin}}}, \bibinfo
  {author} {\bibfnamefont {C.}~\bibnamefont {{Heymans}}}, \bibinfo {author}
  {\bibfnamefont {H.}~\bibnamefont {{Hildebrandt}}}, \ and\ \bibinfo {author}
  {\bibfnamefont {Q.}~\bibnamefont {{Xia}}},\ }\href {\doibase
  10.1093/mnras/stab1623} {\bibfield  {journal} {\bibinfo  {journal} {\mnras}\
  }\textbf {\bibinfo {volume} {506}},\ \bibinfo {pages} {1623} (\bibinfo {year}
  {2021})},\ \Eprint {http://arxiv.org/abs/2012.02777} {arXiv:2012.02777}
  \BibitemShut {NoStop}%
\bibitem [{\citenamefont {{Harnois-Deraps}}\ \emph {et~al.}(2024)\citenamefont
  {{Harnois-Deraps}}, \citenamefont {{Heydenreich}}, \citenamefont {{Giblin}},
  \citenamefont {{Martinet}}, \citenamefont {{Troester}}, \citenamefont
  {{Asgari}}, \citenamefont {{Burger}}, \citenamefont {{Castro}}, \citenamefont
  {{Dolag}}, \citenamefont {{Heymans}} \emph {et~al.}}]{Harnois2024}%
  \BibitemOpen
  \bibfield  {author} {\bibinfo {author} {\bibfnamefont {J.}~\bibnamefont
  {{Harnois-Deraps}}}, \bibinfo {author} {\bibfnamefont {S.}~\bibnamefont
  {{Heydenreich}}}, \bibinfo {author} {\bibfnamefont {B.}~\bibnamefont
  {{Giblin}}}, \bibinfo {author} {\bibfnamefont {N.}~\bibnamefont
  {{Martinet}}}, \bibinfo {author} {\bibfnamefont {T.}~\bibnamefont
  {{Troester}}}, \bibinfo {author} {\bibfnamefont {M.}~\bibnamefont
  {{Asgari}}}, \bibinfo {author} {\bibfnamefont {P.}~\bibnamefont {{Burger}}},
  \bibinfo {author} {\bibfnamefont {T.}~\bibnamefont {{Castro}}}, \bibinfo
  {author} {\bibfnamefont {K.}~\bibnamefont {{Dolag}}}, \bibinfo {author}
  {\bibfnamefont {C.}~\bibnamefont {{Heymans}}},  \emph {et~al.},\ }\href@noop
  {} {\  (\bibinfo {year} {2024})},\ \Eprint {http://arxiv.org/abs/2405.10312}
  {arXiv:2405.10312 [astro-ph.CO]} \BibitemShut {NoStop}%
\bibitem [{\citenamefont {{Heydenreich}}\ \emph {et~al.}(2021)\citenamefont
  {{Heydenreich}}, \citenamefont {{Br{\"u}ck}},\ and\ \citenamefont
  {{Harnois-D{\'e}raps}}}]{Heydenreich:2021}%
  \BibitemOpen
  \bibfield  {author} {\bibinfo {author} {\bibfnamefont {S.}~\bibnamefont
  {{Heydenreich}}}, \bibinfo {author} {\bibfnamefont {B.}~\bibnamefont
  {{Br{\"u}ck}}}, \ and\ \bibinfo {author} {\bibfnamefont {J.}~\bibnamefont
  {{Harnois-D{\'e}raps}}},\ }\href {\doibase 10.1051/0004-6361/202039048}
  {\bibfield  {journal} {\bibinfo  {journal} {\aap}\ }\textbf {\bibinfo
  {volume} {648}},\ \bibinfo {eid} {A74} (\bibinfo {year} {2021})},\ \Eprint
  {http://arxiv.org/abs/2007.13724} {arXiv:2007.13724} \BibitemShut {NoStop}%
\bibitem [{\citenamefont {{Heydenreich}}\ \emph {et~al.}(2022)\citenamefont
  {{Heydenreich}}, \citenamefont {{Br{\"u}ck}}, \citenamefont {{Burger}},
  \citenamefont {{Harnois-D{\'e}raps}}, \citenamefont {{Unruh}}, \citenamefont
  {{Castro}}, \citenamefont {{Dolag}},\ and\ \citenamefont
  {{Martinet}}}]{Heydenreich2022}%
  \BibitemOpen
  \bibfield  {author} {\bibinfo {author} {\bibfnamefont {S.}~\bibnamefont
  {{Heydenreich}}}, \bibinfo {author} {\bibfnamefont {B.}~\bibnamefont
  {{Br{\"u}ck}}}, \bibinfo {author} {\bibfnamefont {P.}~\bibnamefont
  {{Burger}}}, \bibinfo {author} {\bibfnamefont {J.}~\bibnamefont
  {{Harnois-D{\'e}raps}}}, \bibinfo {author} {\bibfnamefont {S.}~\bibnamefont
  {{Unruh}}}, \bibinfo {author} {\bibfnamefont {T.}~\bibnamefont {{Castro}}},
  \bibinfo {author} {\bibfnamefont {K.}~\bibnamefont {{Dolag}}}, \ and\
  \bibinfo {author} {\bibfnamefont {N.}~\bibnamefont {{Martinet}}},\ }\href
  {\doibase 10.1051/0004-6361/202243868} {\bibfield  {journal} {\bibinfo
  {journal} {\aap}\ }\textbf {\bibinfo {volume} {667}},\ \bibinfo {eid} {A125}
  (\bibinfo {year} {2022})},\ \Eprint {http://arxiv.org/abs/2204.11831}
  {arXiv:2204.11831} \BibitemShut {NoStop}%
\bibitem [{\citenamefont {{Halder}}\ \emph {et~al.}(2021)\citenamefont
  {{Halder}}, \citenamefont {{Friedrich}}, \citenamefont {{Seitz}},\ and\
  \citenamefont {{Varga}}}]{Halder:2021}%
  \BibitemOpen
  \bibfield  {author} {\bibinfo {author} {\bibfnamefont {A.}~\bibnamefont
  {{Halder}}}, \bibinfo {author} {\bibfnamefont {O.}~\bibnamefont
  {{Friedrich}}}, \bibinfo {author} {\bibfnamefont {S.}~\bibnamefont
  {{Seitz}}}, \ and\ \bibinfo {author} {\bibfnamefont {T.~N.}\ \bibnamefont
  {{Varga}}},\ }\href {\doibase 10.1093/mnras/stab1801} {\bibfield  {journal}
  {\bibinfo  {journal} {\mnras}\ }\textbf {\bibinfo {volume} {506}},\ \bibinfo
  {pages} {2780} (\bibinfo {year} {2021})},\ \Eprint
  {http://arxiv.org/abs/2102.10177} {arXiv:2102.10177} \BibitemShut {NoStop}%
\bibitem [{\citenamefont {{Halder}}\ and\ \citenamefont
  {{Barreira}}(2022)}]{Halder:2022}%
  \BibitemOpen
  \bibfield  {author} {\bibinfo {author} {\bibfnamefont {A.}~\bibnamefont
  {{Halder}}}\ and\ \bibinfo {author} {\bibfnamefont {A.}~\bibnamefont
  {{Barreira}}},\ }\href {\doibase 10.1093/mnras/stac2046} {\bibfield
  {journal} {\bibinfo  {journal} {\mnras}\ }\textbf {\bibinfo {volume} {515}},\
  \bibinfo {pages} {4639} (\bibinfo {year} {2022})},\ \Eprint
  {http://arxiv.org/abs/2201.05607} {arXiv:2201.05607} \BibitemShut {NoStop}%
\bibitem [{\citenamefont {{Gatti}}\ \emph {et~al.}(2022)\citenamefont
  {{Gatti}}, \citenamefont {{Jain}}, \citenamefont {{Chang}}, \citenamefont
  {{Raveri}}, \citenamefont {{Z{\"u}rcher}}, \citenamefont {{Secco}},
  \citenamefont {{Whiteway}}, \citenamefont {{Jeffrey}}, \citenamefont
  {{Doux}}, \citenamefont {{Kacprzak}} \emph {et~al.}}]{Gatti2022}%
  \BibitemOpen
  \bibfield  {author} {\bibinfo {author} {\bibfnamefont {M.}~\bibnamefont
  {{Gatti}}}, \bibinfo {author} {\bibfnamefont {B.}~\bibnamefont {{Jain}}},
  \bibinfo {author} {\bibfnamefont {C.}~\bibnamefont {{Chang}}}, \bibinfo
  {author} {\bibfnamefont {M.}~\bibnamefont {{Raveri}}}, \bibinfo {author}
  {\bibfnamefont {D.}~\bibnamefont {{Z{\"u}rcher}}}, \bibinfo {author}
  {\bibfnamefont {L.}~\bibnamefont {{Secco}}}, \bibinfo {author} {\bibfnamefont
  {L.}~\bibnamefont {{Whiteway}}}, \bibinfo {author} {\bibfnamefont
  {N.}~\bibnamefont {{Jeffrey}}}, \bibinfo {author} {\bibfnamefont
  {C.}~\bibnamefont {{Doux}}}, \bibinfo {author} {\bibfnamefont
  {T.}~\bibnamefont {{Kacprzak}}},  \emph {et~al.},\ }\href {\doibase
  10.1103/PhysRevD.106.083509} {\bibfield  {journal} {\bibinfo  {journal}
  {\prd}\ }\textbf {\bibinfo {volume} {106}},\ \bibinfo {eid} {083509}
  (\bibinfo {year} {2022})},\ \Eprint {http://arxiv.org/abs/2110.10141}
  {arXiv:2110.10141} \BibitemShut {NoStop}%
\bibitem [{\citenamefont {{Burger}}\ \emph {et~al.}(2024)\citenamefont
  {{Burger}}, \citenamefont {{Porth}}, \citenamefont {{Heydenreich}},
  \citenamefont {{Linke}}, \citenamefont {{Wielders}}, \citenamefont
  {{Schneider}}, \citenamefont {{Asgari}}, \citenamefont {{Castro}},
  \citenamefont {{Dolag}}, \citenamefont {{Harnois-D{\'e}raps}} \emph
  {et~al.}}]{Burger2024}%
  \BibitemOpen
  \bibfield  {author} {\bibinfo {author} {\bibfnamefont {P.~A.}\ \bibnamefont
  {{Burger}}}, \bibinfo {author} {\bibfnamefont {L.}~\bibnamefont {{Porth}}},
  \bibinfo {author} {\bibfnamefont {S.}~\bibnamefont {{Heydenreich}}}, \bibinfo
  {author} {\bibfnamefont {L.}~\bibnamefont {{Linke}}}, \bibinfo {author}
  {\bibfnamefont {N.}~\bibnamefont {{Wielders}}}, \bibinfo {author}
  {\bibfnamefont {P.}~\bibnamefont {{Schneider}}}, \bibinfo {author}
  {\bibfnamefont {M.}~\bibnamefont {{Asgari}}}, \bibinfo {author}
  {\bibfnamefont {T.}~\bibnamefont {{Castro}}}, \bibinfo {author}
  {\bibfnamefont {K.}~\bibnamefont {{Dolag}}}, \bibinfo {author} {\bibfnamefont
  {J.}~\bibnamefont {{Harnois-D{\'e}raps}}},  \emph {et~al.},\ }\href {\doibase
  10.1051/0004-6361/202347986} {\bibfield  {journal} {\bibinfo  {journal}
  {\aap}\ }\textbf {\bibinfo {volume} {683}},\ \bibinfo {eid} {A103} (\bibinfo
  {year} {2024})},\ \Eprint {http://arxiv.org/abs/2309.08602}
  {arXiv:2309.08602} \BibitemShut {NoStop}%
\bibitem [{\citenamefont {{van Uitert}}\ \emph {et~al.}(2018)\citenamefont
  {{van Uitert}}, \citenamefont {{Joachimi}}, \citenamefont {{Joudaki}},
  \citenamefont {{Amon}}, \citenamefont {{Heymans}}, \citenamefont
  {{K{\"o}hlinger}}, \citenamefont {{Asgari}}, \citenamefont {{Blake}},
  \citenamefont {{Choi}}, \citenamefont {{Erben}} \emph
  {et~al.}}]{vanUitert2018}%
  \BibitemOpen
  \bibfield  {author} {\bibinfo {author} {\bibfnamefont {E.}~\bibnamefont {{van
  Uitert}}}, \bibinfo {author} {\bibfnamefont {B.}~\bibnamefont {{Joachimi}}},
  \bibinfo {author} {\bibfnamefont {S.}~\bibnamefont {{Joudaki}}}, \bibinfo
  {author} {\bibfnamefont {A.}~\bibnamefont {{Amon}}}, \bibinfo {author}
  {\bibfnamefont {C.}~\bibnamefont {{Heymans}}}, \bibinfo {author}
  {\bibfnamefont {F.}~\bibnamefont {{K{\"o}hlinger}}}, \bibinfo {author}
  {\bibfnamefont {M.}~\bibnamefont {{Asgari}}}, \bibinfo {author}
  {\bibfnamefont {C.}~\bibnamefont {{Blake}}}, \bibinfo {author} {\bibfnamefont
  {A.}~\bibnamefont {{Choi}}}, \bibinfo {author} {\bibfnamefont
  {T.}~\bibnamefont {{Erben}}},  \emph {et~al.},\ }\href {\doibase
  10.1093/mnras/sty551} {\bibfield  {journal} {\bibinfo  {journal} {\mnras}\
  }\textbf {\bibinfo {volume} {476}},\ \bibinfo {pages} {4662} (\bibinfo {year}
  {2018})},\ \Eprint {http://arxiv.org/abs/1706.05004} {arXiv:1706.05004}
  \BibitemShut {NoStop}%
\bibitem [{\citenamefont {{Joudaki}}\ \emph {et~al.}(2018)\citenamefont
  {{Joudaki}}, \citenamefont {{Blake}}, \citenamefont {{Johnson}},
  \citenamefont {{Amon}}, \citenamefont {{Asgari}}, \citenamefont {{Choi}},
  \citenamefont {{Erben}}, \citenamefont {{Glazebrook}}, \citenamefont
  {{Harnois-D{\'e}raps}} \emph {et~al.}}]{Joudaki2018}%
  \BibitemOpen
  \bibfield  {author} {\bibinfo {author} {\bibfnamefont {S.}~\bibnamefont
  {{Joudaki}}}, \bibinfo {author} {\bibfnamefont {C.}~\bibnamefont {{Blake}}},
  \bibinfo {author} {\bibfnamefont {A.}~\bibnamefont {{Johnson}}}, \bibinfo
  {author} {\bibfnamefont {A.}~\bibnamefont {{Amon}}}, \bibinfo {author}
  {\bibfnamefont {M.}~\bibnamefont {{Asgari}}}, \bibinfo {author}
  {\bibfnamefont {A.}~\bibnamefont {{Choi}}}, \bibinfo {author} {\bibfnamefont
  {T.}~\bibnamefont {{Erben}}}, \bibinfo {author} {\bibfnamefont
  {K.}~\bibnamefont {{Glazebrook}}}, \bibinfo {author} {\bibfnamefont
  {J.}~\bibnamefont {{Harnois-D{\'e}raps}}},  \emph {et~al.},\ }\href {\doibase
  10.1093/mnras/stx2820} {\bibfield  {journal} {\bibinfo  {journal} {\mnras}\
  }\textbf {\bibinfo {volume} {474}},\ \bibinfo {pages} {4894} (\bibinfo {year}
  {2018})},\ \Eprint {http://arxiv.org/abs/1707.06627} {arXiv:1707.06627}
  \BibitemShut {NoStop}%
\bibitem [{\citenamefont {{DES Collaboration}}\ \emph
  {et~al.}(2018)\citenamefont {{DES Collaboration}}, \citenamefont {{Abbott}},
  \citenamefont {{Abdalla}}, \citenamefont {{Alarcon}}, \citenamefont
  {{Aleksi{\'c}}}, \citenamefont {{Allam}}, \citenamefont {{Allen}},
  \citenamefont {{Amara}}, \citenamefont {{Annis}}, \citenamefont {{Asorey}}
  \emph {et~al.}}]{DES:2018}%
  \BibitemOpen
  \bibfield  {author} {\bibinfo {author} {\bibnamefont {{DES Collaboration}}},
  \bibinfo {author} {\bibfnamefont {T.~M.~C.}\ \bibnamefont {{Abbott}}},
  \bibinfo {author} {\bibfnamefont {F.~B.}\ \bibnamefont {{Abdalla}}}, \bibinfo
  {author} {\bibfnamefont {A.}~\bibnamefont {{Alarcon}}}, \bibinfo {author}
  {\bibfnamefont {J.}~\bibnamefont {{Aleksi{\'c}}}}, \bibinfo {author}
  {\bibfnamefont {S.}~\bibnamefont {{Allam}}}, \bibinfo {author} {\bibfnamefont
  {S.}~\bibnamefont {{Allen}}}, \bibinfo {author} {\bibfnamefont
  {A.}~\bibnamefont {{Amara}}}, \bibinfo {author} {\bibfnamefont
  {J.}~\bibnamefont {{Annis}}}, \bibinfo {author} {\bibfnamefont
  {J.}~\bibnamefont {{Asorey}}},  \emph {et~al.},\ }\href {\doibase
  10.1103/PhysRevD.98.043526} {\bibfield  {journal} {\bibinfo  {journal}
  {\prd}\ }\textbf {\bibinfo {volume} {98}},\ \bibinfo {eid} {043526} (\bibinfo
  {year} {2018})},\ \Eprint {http://arxiv.org/abs/1708.01530}
  {arXiv:1708.01530} \BibitemShut {NoStop}%
\bibitem [{\citenamefont {{DES Collaboration}}\ \emph
  {et~al.}(2022)\citenamefont {{DES Collaboration}}, \citenamefont {{Abbott}},
  \citenamefont {{Aguena}}, \citenamefont {{Alarcon}}, \citenamefont {{Allam}},
  \citenamefont {{Alves}}, \citenamefont {{Amon}}, \citenamefont
  {{Andrade-Oliveira}}, \citenamefont {{Annis}}, \citenamefont {{Avila}} \emph
  {et~al.}}]{DES2021}%
  \BibitemOpen
  \bibfield  {author} {\bibinfo {author} {\bibnamefont {{DES Collaboration}}},
  \bibinfo {author} {\bibfnamefont {T.~M.~C.}\ \bibnamefont {{Abbott}}},
  \bibinfo {author} {\bibfnamefont {M.}~\bibnamefont {{Aguena}}}, \bibinfo
  {author} {\bibfnamefont {A.}~\bibnamefont {{Alarcon}}}, \bibinfo {author}
  {\bibfnamefont {S.}~\bibnamefont {{Allam}}}, \bibinfo {author} {\bibfnamefont
  {O.}~\bibnamefont {{Alves}}}, \bibinfo {author} {\bibfnamefont
  {A.}~\bibnamefont {{Amon}}}, \bibinfo {author} {\bibfnamefont
  {F.}~\bibnamefont {{Andrade-Oliveira}}}, \bibinfo {author} {\bibfnamefont
  {J.}~\bibnamefont {{Annis}}}, \bibinfo {author} {\bibfnamefont
  {S.}~\bibnamefont {{Avila}}},  \emph {et~al.},\ }\href {\doibase
  10.1103/PhysRevD.105.023520} {\bibfield  {journal} {\bibinfo  {journal}
  {\prd}\ }\textbf {\bibinfo {volume} {105}},\ \bibinfo {eid} {023520}
  (\bibinfo {year} {2022})},\ \Eprint {http://arxiv.org/abs/2105.13549}
  {arXiv:2105.13549} \BibitemShut {NoStop}%
\bibitem [{\citenamefont {{Gruen}}\ \emph {et~al.}(2018)\citenamefont
  {{Gruen}}, \citenamefont {{Friedrich}}, \citenamefont {{Krause}},
  \citenamefont {{DeRose}}, \citenamefont {{Cawthon}}, \citenamefont {{Davis}},
  \citenamefont {{Elvin-Poole}}, \citenamefont {{Rykoff}}, \citenamefont
  {{Wechsler}}, \citenamefont {{Alarcon}} \emph {et~al.}}]{Gruen:2018}%
  \BibitemOpen
  \bibfield  {author} {\bibinfo {author} {\bibfnamefont {D.}~\bibnamefont
  {{Gruen}}}, \bibinfo {author} {\bibfnamefont {O.}~\bibnamefont
  {{Friedrich}}}, \bibinfo {author} {\bibfnamefont {E.}~\bibnamefont
  {{Krause}}}, \bibinfo {author} {\bibfnamefont {J.}~\bibnamefont {{DeRose}}},
  \bibinfo {author} {\bibfnamefont {R.}~\bibnamefont {{Cawthon}}}, \bibinfo
  {author} {\bibfnamefont {C.}~\bibnamefont {{Davis}}}, \bibinfo {author}
  {\bibfnamefont {J.}~\bibnamefont {{Elvin-Poole}}}, \bibinfo {author}
  {\bibfnamefont {E.~S.}\ \bibnamefont {{Rykoff}}}, \bibinfo {author}
  {\bibfnamefont {R.~H.}\ \bibnamefont {{Wechsler}}}, \bibinfo {author}
  {\bibfnamefont {A.}~\bibnamefont {{Alarcon}}},  \emph {et~al.},\ }\href
  {\doibase 10.1103/PhysRevD.98.023507} {\bibfield  {journal} {\bibinfo
  {journal} {\prd}\ }\textbf {\bibinfo {volume} {98}},\ \bibinfo {eid} {023507}
  (\bibinfo {year} {2018})},\ \Eprint {http://arxiv.org/abs/1710.05045}
  {arXiv:1710.05045} \BibitemShut {NoStop}%
\bibitem [{\citenamefont {{Burger}}\ \emph {et~al.}(2023)\citenamefont
  {{Burger}}, \citenamefont {{Friedrich}}, \citenamefont
  {{Harnois-D{\'e}raps}}, \citenamefont {{Schneider}}, \citenamefont
  {{Asgari}}, \citenamefont {{Bilicki}}, \citenamefont {{Hildebrandt}},
  \citenamefont {{Wright}}, \citenamefont {{Castro}}, \citenamefont {{Dolag}}
  \emph {et~al.}}]{Burger2023}%
  \BibitemOpen
  \bibfield  {author} {\bibinfo {author} {\bibfnamefont {P.~A.}\ \bibnamefont
  {{Burger}}}, \bibinfo {author} {\bibfnamefont {O.}~\bibnamefont
  {{Friedrich}}}, \bibinfo {author} {\bibfnamefont {J.}~\bibnamefont
  {{Harnois-D{\'e}raps}}}, \bibinfo {author} {\bibfnamefont {P.}~\bibnamefont
  {{Schneider}}}, \bibinfo {author} {\bibfnamefont {M.}~\bibnamefont
  {{Asgari}}}, \bibinfo {author} {\bibfnamefont {M.}~\bibnamefont {{Bilicki}}},
  \bibinfo {author} {\bibfnamefont {H.}~\bibnamefont {{Hildebrandt}}}, \bibinfo
  {author} {\bibfnamefont {A.~H.}\ \bibnamefont {{Wright}}}, \bibinfo {author}
  {\bibfnamefont {T.}~\bibnamefont {{Castro}}}, \bibinfo {author}
  {\bibfnamefont {K.}~\bibnamefont {{Dolag}}},  \emph {et~al.},\ }\href
  {\doibase 10.1051/0004-6361/202244673} {\bibfield  {journal} {\bibinfo
  {journal} {\aap}\ }\textbf {\bibinfo {volume} {669}},\ \bibinfo {eid} {A69}
  (\bibinfo {year} {2023})},\ \Eprint {http://arxiv.org/abs/2208.02171}
  {arXiv:2208.02171} \BibitemShut {NoStop}%
\bibitem [{\citenamefont {{Paillas}}\ \emph {et~al.}(2023)\citenamefont
  {{Paillas}}, \citenamefont {{Cuesta-Lazaro}}, \citenamefont {{Percival}},
  \citenamefont {{Nadathur}}, \citenamefont {{Cai}}, \citenamefont {{Yuan}},
  \citenamefont {{Beutler}}, \citenamefont {{de Mattia}}, \citenamefont
  {{Eisenstein}}, \citenamefont {{Forero-Sanchez}} \emph
  {et~al.}}]{Paillas2023}%
  \BibitemOpen
  \bibfield  {author} {\bibinfo {author} {\bibfnamefont {E.}~\bibnamefont
  {{Paillas}}}, \bibinfo {author} {\bibfnamefont {C.}~\bibnamefont
  {{Cuesta-Lazaro}}}, \bibinfo {author} {\bibfnamefont {W.~J.}\ \bibnamefont
  {{Percival}}}, \bibinfo {author} {\bibfnamefont {S.}~\bibnamefont
  {{Nadathur}}}, \bibinfo {author} {\bibfnamefont {Y.-C.}\ \bibnamefont
  {{Cai}}}, \bibinfo {author} {\bibfnamefont {S.}~\bibnamefont {{Yuan}}},
  \bibinfo {author} {\bibfnamefont {F.}~\bibnamefont {{Beutler}}}, \bibinfo
  {author} {\bibfnamefont {A.}~\bibnamefont {{de Mattia}}}, \bibinfo {author}
  {\bibfnamefont {D.}~\bibnamefont {{Eisenstein}}}, \bibinfo {author}
  {\bibfnamefont {D.}~\bibnamefont {{Forero-Sanchez}}},  \emph {et~al.},\
  }\href@noop {} {\  (\bibinfo {year} {2023})},\ \Eprint
  {http://arxiv.org/abs/2309.16541} {arXiv:2309.16541} \BibitemShut {NoStop}%
\bibitem [{\citenamefont {{Friedrich}}\ \emph {et~al.}(2018)\citenamefont
  {{Friedrich}}, \citenamefont {{Gruen}}, \citenamefont {{DeRose}},
  \citenamefont {{Kirk}}, \citenamefont {{Krause}}, \citenamefont
  {{McClintock}}, \citenamefont {{Rykoff}}, \citenamefont {{Seitz}},
  \citenamefont {{Wechsler}}, \citenamefont {{Bernstein}} \emph
  {et~al.}}]{Friedrich:2018}%
  \BibitemOpen
  \bibfield  {author} {\bibinfo {author} {\bibfnamefont {O.}~\bibnamefont
  {{Friedrich}}}, \bibinfo {author} {\bibfnamefont {D.}~\bibnamefont
  {{Gruen}}}, \bibinfo {author} {\bibfnamefont {J.}~\bibnamefont {{DeRose}}},
  \bibinfo {author} {\bibfnamefont {D.}~\bibnamefont {{Kirk}}}, \bibinfo
  {author} {\bibfnamefont {E.}~\bibnamefont {{Krause}}}, \bibinfo {author}
  {\bibfnamefont {T.}~\bibnamefont {{McClintock}}}, \bibinfo {author}
  {\bibfnamefont {E.~S.}\ \bibnamefont {{Rykoff}}}, \bibinfo {author}
  {\bibfnamefont {S.}~\bibnamefont {{Seitz}}}, \bibinfo {author} {\bibfnamefont
  {R.~H.}\ \bibnamefont {{Wechsler}}}, \bibinfo {author} {\bibfnamefont
  {G.~M.}\ \bibnamefont {{Bernstein}}},  \emph {et~al.},\ }\href {\doibase
  10.1103/PhysRevD.98.023508} {\bibfield  {journal} {\bibinfo  {journal}
  {\prd}\ }\textbf {\bibinfo {volume} {98}},\ \bibinfo {eid} {023508} (\bibinfo
  {year} {2018})},\ \Eprint {http://arxiv.org/abs/1710.05162}
  {arXiv:1710.05162} \BibitemShut {NoStop}%
\bibitem [{\citenamefont {{Burger}}\ \emph {et~al.}(2022)\citenamefont
  {{Burger}}, \citenamefont {{Friedrich}}, \citenamefont
  {{Harnois-D{\'e}raps}},\ and\ \citenamefont {{Schneider}}}]{Burger2022}%
  \BibitemOpen
  \bibfield  {author} {\bibinfo {author} {\bibfnamefont {P.}~\bibnamefont
  {{Burger}}}, \bibinfo {author} {\bibfnamefont {O.}~\bibnamefont
  {{Friedrich}}}, \bibinfo {author} {\bibfnamefont {J.}~\bibnamefont
  {{Harnois-D{\'e}raps}}}, \ and\ \bibinfo {author} {\bibfnamefont
  {P.}~\bibnamefont {{Schneider}}},\ }\href {\doibase
  10.1051/0004-6361/202141628} {\bibfield  {journal} {\bibinfo  {journal}
  {\aap}\ }\textbf {\bibinfo {volume} {661}},\ \bibinfo {eid} {A137} (\bibinfo
  {year} {2022})},\ \Eprint {http://arxiv.org/abs/2106.13214}
  {arXiv:2106.13214} \BibitemShut {NoStop}%
\bibitem [{\citenamefont {{Dawson}}\ \emph {et~al.}(2013)\citenamefont
  {{Dawson}}, \citenamefont {{Schlegel}}, \citenamefont {{Ahn}}, \citenamefont
  {{Anderson}}, \citenamefont {{Aubourg}}, \citenamefont {{Bailey}},
  \citenamefont {{Barkhouser}}, \citenamefont {{Bautista}}, \citenamefont
  {{Beifiori}}, \citenamefont {{Berlind}} \emph {et~al.}}]{Dawson2013}%
  \BibitemOpen
  \bibfield  {author} {\bibinfo {author} {\bibfnamefont {K.~S.}\ \bibnamefont
  {{Dawson}}}, \bibinfo {author} {\bibfnamefont {D.~J.}\ \bibnamefont
  {{Schlegel}}}, \bibinfo {author} {\bibfnamefont {C.~P.}\ \bibnamefont
  {{Ahn}}}, \bibinfo {author} {\bibfnamefont {S.~F.}\ \bibnamefont
  {{Anderson}}}, \bibinfo {author} {\bibfnamefont {{\'E}.}~\bibnamefont
  {{Aubourg}}}, \bibinfo {author} {\bibfnamefont {S.}~\bibnamefont {{Bailey}}},
  \bibinfo {author} {\bibfnamefont {R.~H.}\ \bibnamefont {{Barkhouser}}},
  \bibinfo {author} {\bibfnamefont {J.~E.}\ \bibnamefont {{Bautista}}},
  \bibinfo {author} {\bibfnamefont {A.}~\bibnamefont {{Beifiori}}}, \bibinfo
  {author} {\bibfnamefont {A.~A.}\ \bibnamefont {{Berlind}}},  \emph {et~al.},\
  }\href {\doibase 10.1088/0004-6256/145/1/10} {\bibfield  {journal} {\bibinfo
  {journal} {\aj}\ }\textbf {\bibinfo {volume} {145}},\ \bibinfo {eid} {10}
  (\bibinfo {year} {2013})},\ \Eprint {http://arxiv.org/abs/1208.0022}
  {arXiv:1208.0022} \BibitemShut {NoStop}%
\bibitem [{\citenamefont {{Cuesta-Lazaro}}\ \emph {et~al.}(2023)\citenamefont
  {{Cuesta-Lazaro}}, \citenamefont {{Paillas}}, \citenamefont {{Yuan}},
  \citenamefont {{Cai}}, \citenamefont {{Nadathur}}, \citenamefont
  {{Percival}}, \citenamefont {{Beutler}}, \citenamefont {{de Mattia}},
  \citenamefont {{Eisenstein}}, \citenamefont {{Forero-Sanchez}} \emph
  {et~al.}}]{Cuesta-Lazaro2023}%
  \BibitemOpen
  \bibfield  {author} {\bibinfo {author} {\bibfnamefont {C.}~\bibnamefont
  {{Cuesta-Lazaro}}}, \bibinfo {author} {\bibfnamefont {E.}~\bibnamefont
  {{Paillas}}}, \bibinfo {author} {\bibfnamefont {S.}~\bibnamefont {{Yuan}}},
  \bibinfo {author} {\bibfnamefont {Y.-C.}\ \bibnamefont {{Cai}}}, \bibinfo
  {author} {\bibfnamefont {S.}~\bibnamefont {{Nadathur}}}, \bibinfo {author}
  {\bibfnamefont {W.~J.}\ \bibnamefont {{Percival}}}, \bibinfo {author}
  {\bibfnamefont {F.}~\bibnamefont {{Beutler}}}, \bibinfo {author}
  {\bibfnamefont {A.}~\bibnamefont {{de Mattia}}}, \bibinfo {author}
  {\bibfnamefont {D.}~\bibnamefont {{Eisenstein}}}, \bibinfo {author}
  {\bibfnamefont {D.}~\bibnamefont {{Forero-Sanchez}}},  \emph {et~al.},\
  }\href@noop {} {\  (\bibinfo {year} {2023})},\ \Eprint
  {http://arxiv.org/abs/2309.16539} {arXiv:2309.16539} \BibitemShut {NoStop}%
\bibitem [{\citenamefont {{Maksimova}}\ \emph {et~al.}(2021)\citenamefont
  {{Maksimova}}, \citenamefont {{Garrison}}, \citenamefont {{Eisenstein}},
  \citenamefont {{Hadzhiyska}}, \citenamefont {{Bose}},\ and\ \citenamefont
  {{Satterthwaite}}}]{Maksimova2021}%
  \BibitemOpen
  \bibfield  {author} {\bibinfo {author} {\bibfnamefont {N.~A.}\ \bibnamefont
  {{Maksimova}}}, \bibinfo {author} {\bibfnamefont {L.~H.}\ \bibnamefont
  {{Garrison}}}, \bibinfo {author} {\bibfnamefont {D.~J.}\ \bibnamefont
  {{Eisenstein}}}, \bibinfo {author} {\bibfnamefont {B.}~\bibnamefont
  {{Hadzhiyska}}}, \bibinfo {author} {\bibfnamefont {S.}~\bibnamefont
  {{Bose}}}, \ and\ \bibinfo {author} {\bibfnamefont {T.~P.}\ \bibnamefont
  {{Satterthwaite}}},\ }\href {\doibase 10.1093/mnras/stab2484} {\bibfield
  {journal} {\bibinfo  {journal} {\mnras}\ }\textbf {\bibinfo {volume} {508}},\
  \bibinfo {pages} {4017} (\bibinfo {year} {2021})},\ \Eprint
  {http://arxiv.org/abs/2110.11398} {arXiv:2110.11398} \BibitemShut {NoStop}%
\bibitem [{\citenamefont {{Seljak}}(2000)}]{Seljak2000}%
  \BibitemOpen
  \bibfield  {author} {\bibinfo {author} {\bibfnamefont {U.}~\bibnamefont
  {{Seljak}}},\ }\href {\doibase 10.1046/j.1365-8711.2000.03715.x} {\bibfield
  {journal} {\bibinfo  {journal} {\mnras}\ }\textbf {\bibinfo {volume} {318}},\
  \bibinfo {pages} {203} (\bibinfo {year} {2000})},\ \Eprint
  {http://arxiv.org/abs/astro-ph/0001493} {arXiv:astro-ph/0001493 [astro-ph]}
  \BibitemShut {NoStop}%
\bibitem [{\citenamefont {{Cooray}}\ and\ \citenamefont
  {{Sheth}}(2002)}]{Cooray2002}%
  \BibitemOpen
  \bibfield  {author} {\bibinfo {author} {\bibfnamefont {A.}~\bibnamefont
  {{Cooray}}}\ and\ \bibinfo {author} {\bibfnamefont {R.}~\bibnamefont
  {{Sheth}}},\ }\href {\doibase 10.1016/S0370-1573(02)00276-4} {\bibfield
  {journal} {\bibinfo  {journal} {Physics Reports}\ }\textbf {\bibinfo {volume}
  {372}},\ \bibinfo {pages} {1} (\bibinfo {year} {2002})},\ \Eprint
  {http://arxiv.org/abs/astro-ph/0206508} {arXiv:astro-ph/0206508} \BibitemShut
  {NoStop}%
\bibitem [{\citenamefont {{Peacock}}\ and\ \citenamefont
  {{Smith}}(2000)}]{Peacock2000}%
  \BibitemOpen
  \bibfield  {author} {\bibinfo {author} {\bibfnamefont {J.~A.}\ \bibnamefont
  {{Peacock}}}\ and\ \bibinfo {author} {\bibfnamefont {R.~E.}\ \bibnamefont
  {{Smith}}},\ }\href {\doibase 10.1046/j.1365-8711.2000.03779.x} {\bibfield
  {journal} {\bibinfo  {journal} {\mnras}\ }\textbf {\bibinfo {volume} {318}},\
  \bibinfo {pages} {1144} (\bibinfo {year} {2000})},\ \Eprint
  {http://arxiv.org/abs/astro-ph/0005010} {arXiv:astro-ph/0005010 [astro-ph]}
  \BibitemShut {NoStop}%
\bibitem [{\citenamefont {{Spurio Mancini}}\ \emph {et~al.}(2022)\citenamefont
  {{Spurio Mancini}}, \citenamefont {{Piras}}, \citenamefont {{Alsing}},
  \citenamefont {{Joachimi}},\ and\ \citenamefont {{Hobson}}}]{COSMOPOWER2022}%
  \BibitemOpen
  \bibfield  {author} {\bibinfo {author} {\bibfnamefont {A.}~\bibnamefont
  {{Spurio Mancini}}}, \bibinfo {author} {\bibfnamefont {D.}~\bibnamefont
  {{Piras}}}, \bibinfo {author} {\bibfnamefont {J.}~\bibnamefont {{Alsing}}},
  \bibinfo {author} {\bibfnamefont {B.}~\bibnamefont {{Joachimi}}}, \ and\
  \bibinfo {author} {\bibfnamefont {M.~P.}\ \bibnamefont {{Hobson}}},\ }\href
  {\doibase 10.1093/mnras/stac064} {\bibfield  {journal} {\bibinfo  {journal}
  {\mnras}\ }\textbf {\bibinfo {volume} {511}},\ \bibinfo {pages} {1771}
  (\bibinfo {year} {2022})},\ \Eprint {http://arxiv.org/abs/2106.03846}
  {arXiv:2106.03846} \BibitemShut {NoStop}%
\bibitem [{\citenamefont {{Euclid Collaboration}}\ \emph
  {et~al.}(2021)\citenamefont {{Euclid Collaboration}}, \citenamefont
  {{Knabenhans}}, \citenamefont {{Stadel}}, \citenamefont {{Potter}},
  \citenamefont {{Dakin}}, \citenamefont {{Hannestad}}, \citenamefont {{Tram}},
  \citenamefont {{Marelli}}, \citenamefont {{Schneider}}, \citenamefont
  {{Teyssier}} \emph {et~al.}}]{EE2021}%
  \BibitemOpen
  \bibfield  {author} {\bibinfo {author} {\bibnamefont {{Euclid
  Collaboration}}}, \bibinfo {author} {\bibfnamefont {M.}~\bibnamefont
  {{Knabenhans}}}, \bibinfo {author} {\bibfnamefont {J.}~\bibnamefont
  {{Stadel}}}, \bibinfo {author} {\bibfnamefont {D.}~\bibnamefont {{Potter}}},
  \bibinfo {author} {\bibfnamefont {J.}~\bibnamefont {{Dakin}}}, \bibinfo
  {author} {\bibfnamefont {S.}~\bibnamefont {{Hannestad}}}, \bibinfo {author}
  {\bibfnamefont {T.}~\bibnamefont {{Tram}}}, \bibinfo {author} {\bibfnamefont
  {S.}~\bibnamefont {{Marelli}}}, \bibinfo {author} {\bibfnamefont
  {A.}~\bibnamefont {{Schneider}}}, \bibinfo {author} {\bibfnamefont
  {R.}~\bibnamefont {{Teyssier}}},  \emph {et~al.},\ }\href {\doibase
  10.1093/mnras/stab1366} {\bibfield  {journal} {\bibinfo  {journal} {\mnras}\
  }\textbf {\bibinfo {volume} {505}},\ \bibinfo {pages} {2840} (\bibinfo {year}
  {2021})},\ \Eprint {http://arxiv.org/abs/2010.11288} {arXiv:2010.11288}
  \BibitemShut {NoStop}%
\bibitem [{\citenamefont {{Angulo}}\ \emph {et~al.}(2021)\citenamefont
  {{Angulo}}, \citenamefont {{Zennaro}}, \citenamefont {{Contreras}},
  \citenamefont {{Aric{\`o}}}, \citenamefont {{Pellejero-Iba{\~n}ez}},\ and\
  \citenamefont {{St{\"u}cker}}}]{Angulo2021}%
  \BibitemOpen
  \bibfield  {author} {\bibinfo {author} {\bibfnamefont {R.~E.}\ \bibnamefont
  {{Angulo}}}, \bibinfo {author} {\bibfnamefont {M.}~\bibnamefont {{Zennaro}}},
  \bibinfo {author} {\bibfnamefont {S.}~\bibnamefont {{Contreras}}}, \bibinfo
  {author} {\bibfnamefont {G.}~\bibnamefont {{Aric{\`o}}}}, \bibinfo {author}
  {\bibfnamefont {M.}~\bibnamefont {{Pellejero-Iba{\~n}ez}}}, \ and\ \bibinfo
  {author} {\bibfnamefont {J.}~\bibnamefont {{St{\"u}cker}}},\ }\href {\doibase
  10.1093/mnras/stab2018} {\bibfield  {journal} {\bibinfo  {journal} {\mnras}\
  }\textbf {\bibinfo {volume} {507}},\ \bibinfo {pages} {5869} (\bibinfo {year}
  {2021})},\ \Eprint {http://arxiv.org/abs/2004.06245} {arXiv:2004.06245
  [astro-ph.CO]} \BibitemShut {NoStop}%
\bibitem [{\citenamefont {{Bonici}}\ \emph {et~al.}(2024)\citenamefont
  {{Bonici}}, \citenamefont {{Baxter}}, \citenamefont {{Bianchini}},\ and\
  \citenamefont {{Ruiz-Zapatero}}}]{Bonici2024}%
  \BibitemOpen
  \bibfield  {author} {\bibinfo {author} {\bibfnamefont {M.}~\bibnamefont
  {{Bonici}}}, \bibinfo {author} {\bibfnamefont {E.}~\bibnamefont {{Baxter}}},
  \bibinfo {author} {\bibfnamefont {F.}~\bibnamefont {{Bianchini}}}, \ and\
  \bibinfo {author} {\bibfnamefont {J.}~\bibnamefont {{Ruiz-Zapatero}}},\
  }\href {\doibase 10.21105/astro.2307.14339} {\bibfield  {journal} {\bibinfo
  {journal} {The Open Journal of Astrophysics}\ }\textbf {\bibinfo {volume}
  {7}},\ \bibinfo {eid} {10} (\bibinfo {year} {2024})},\ \Eprint
  {http://arxiv.org/abs/2307.14339} {arXiv:2307.14339} \BibitemShut {NoStop}%
\bibitem [{\citenamefont {{Schneider}}(1996)}]{Schneider:1996}%
  \BibitemOpen
  \bibfield  {author} {\bibinfo {author} {\bibfnamefont {P.}~\bibnamefont
  {{Schneider}}},\ }\href {\doibase 10.1093/mnras/283.3.837} {\bibfield
  {journal} {\bibinfo  {journal} {\mnras}\ }\textbf {\bibinfo {volume} {283}},\
  \bibinfo {pages} {837} (\bibinfo {year} {1996})},\ \Eprint
  {http://arxiv.org/abs/astro-ph/9601039} {arXiv:astro-ph/9601039} \BibitemShut
  {NoStop}%
\bibitem [{\citenamefont {{Landy}}\ and\ \citenamefont
  {{Szalay}}(1993)}]{Landy1993}%
  \BibitemOpen
  \bibfield  {author} {\bibinfo {author} {\bibfnamefont {S.~D.}\ \bibnamefont
  {{Landy}}}\ and\ \bibinfo {author} {\bibfnamefont {A.~S.}\ \bibnamefont
  {{Szalay}}},\ }\href {\doibase 10.1086/172900} {\bibfield  {journal}
  {\bibinfo  {journal} {\apj}\ }\textbf {\bibinfo {volume} {412}},\ \bibinfo
  {pages} {64} (\bibinfo {year} {1993})}\BibitemShut {NoStop}%
\bibitem [{\citenamefont {{Schneider}}\ \emph {et~al.}(1998)\citenamefont
  {{Schneider}}, \citenamefont {{van Waerbeke}}, \citenamefont {{Jain}},\ and\
  \citenamefont {{Kruse}}}]{Schneider:1998}%
  \BibitemOpen
  \bibfield  {author} {\bibinfo {author} {\bibfnamefont {P.}~\bibnamefont
  {{Schneider}}}, \bibinfo {author} {\bibfnamefont {L.}~\bibnamefont {{van
  Waerbeke}}}, \bibinfo {author} {\bibfnamefont {B.}~\bibnamefont {{Jain}}}, \
  and\ \bibinfo {author} {\bibfnamefont {G.}~\bibnamefont {{Kruse}}},\ }\href
  {\doibase 10.1046/j.1365-8711.1998.01422.x} {\bibfield  {journal} {\bibinfo
  {journal} {\mnras}\ }\textbf {\bibinfo {volume} {296}},\ \bibinfo {pages}
  {873} (\bibinfo {year} {1998})},\ \Eprint
  {http://arxiv.org/abs/astro-ph/9708143} {arXiv:astro-ph/9708143 [astro-ph]}
  \BibitemShut {NoStop}%
\bibitem [{\citenamefont {{Gruen}}\ \emph {et~al.}(2016)\citenamefont
  {{Gruen}}, \citenamefont {{Friedrich}}, \citenamefont {{Amara}},
  \citenamefont {{Bacon}}, \citenamefont {{Bonnett}}, \citenamefont
  {{Hartley}}, \citenamefont {{Jain}}, \citenamefont {{Jarvis}}, \citenamefont
  {{Kacprzak}}, \citenamefont {{Krause}} \emph {et~al.}}]{Gruen:2015}%
  \BibitemOpen
  \bibfield  {author} {\bibinfo {author} {\bibfnamefont {D.}~\bibnamefont
  {{Gruen}}}, \bibinfo {author} {\bibfnamefont {O.}~\bibnamefont
  {{Friedrich}}}, \bibinfo {author} {\bibfnamefont {A.}~\bibnamefont
  {{Amara}}}, \bibinfo {author} {\bibfnamefont {D.}~\bibnamefont {{Bacon}}},
  \bibinfo {author} {\bibfnamefont {C.}~\bibnamefont {{Bonnett}}}, \bibinfo
  {author} {\bibfnamefont {W.}~\bibnamefont {{Hartley}}}, \bibinfo {author}
  {\bibfnamefont {B.}~\bibnamefont {{Jain}}}, \bibinfo {author} {\bibfnamefont
  {M.}~\bibnamefont {{Jarvis}}}, \bibinfo {author} {\bibfnamefont
  {T.}~\bibnamefont {{Kacprzak}}}, \bibinfo {author} {\bibfnamefont
  {E.}~\bibnamefont {{Krause}}},  \emph {et~al.},\ }\href {\doibase
  10.1093/mnras/stv2506} {\bibfield  {journal} {\bibinfo  {journal} {\mnras}\
  }\textbf {\bibinfo {volume} {455}},\ \bibinfo {pages} {3367} (\bibinfo {year}
  {2016})},\ \Eprint {http://arxiv.org/abs/1507.05090} {arXiv:1507.05090}
  \BibitemShut {NoStop}%
\bibitem [{\citenamefont {{G{\'o}rski}}\ \emph {et~al.}(2005)\citenamefont
  {{G{\'o}rski}}, \citenamefont {{Hivon}}, \citenamefont {{Banday}},
  \citenamefont {{Wandelt}}, \citenamefont {{Hansen}}, \citenamefont
  {{Reinecke}},\ and\ \citenamefont {{Bartelmann}}}]{HEALPix2005}%
  \BibitemOpen
  \bibfield  {author} {\bibinfo {author} {\bibfnamefont {K.~M.}\ \bibnamefont
  {{G{\'o}rski}}}, \bibinfo {author} {\bibfnamefont {E.}~\bibnamefont
  {{Hivon}}}, \bibinfo {author} {\bibfnamefont {A.~J.}\ \bibnamefont
  {{Banday}}}, \bibinfo {author} {\bibfnamefont {B.~D.}\ \bibnamefont
  {{Wandelt}}}, \bibinfo {author} {\bibfnamefont {F.~K.}\ \bibnamefont
  {{Hansen}}}, \bibinfo {author} {\bibfnamefont {M.}~\bibnamefont
  {{Reinecke}}}, \ and\ \bibinfo {author} {\bibfnamefont {M.}~\bibnamefont
  {{Bartelmann}}},\ }\href {\doibase 10.1086/427976} {\bibfield  {journal}
  {\bibinfo  {journal} {\apj}\ }\textbf {\bibinfo {volume} {622}},\ \bibinfo
  {pages} {759} (\bibinfo {year} {2005})},\ \Eprint
  {http://arxiv.org/abs/astro-ph/0409513} {arXiv:astro-ph/0409513 [astro-ph]}
  \BibitemShut {NoStop}%
\bibitem [{\citenamefont {{Guinot}}\ \emph {et~al.}(2022)\citenamefont
  {{Guinot}}, \citenamefont {{Kilbinger}}, \citenamefont {{Farrens}},
  \citenamefont {{Peel}}, \citenamefont {{Pujol}}, \citenamefont {{Schmitz}},
  \citenamefont {{Starck}}, \citenamefont {{Erben}}, \citenamefont {{Gavazzi}},
  \citenamefont {{Gwyn}} \emph {et~al.}}]{Guinot2022}%
  \BibitemOpen
  \bibfield  {author} {\bibinfo {author} {\bibfnamefont {A.}~\bibnamefont
  {{Guinot}}}, \bibinfo {author} {\bibfnamefont {M.}~\bibnamefont
  {{Kilbinger}}}, \bibinfo {author} {\bibfnamefont {S.}~\bibnamefont
  {{Farrens}}}, \bibinfo {author} {\bibfnamefont {A.}~\bibnamefont {{Peel}}},
  \bibinfo {author} {\bibfnamefont {A.}~\bibnamefont {{Pujol}}}, \bibinfo
  {author} {\bibfnamefont {M.}~\bibnamefont {{Schmitz}}}, \bibinfo {author}
  {\bibfnamefont {J.-L.}\ \bibnamefont {{Starck}}}, \bibinfo {author}
  {\bibfnamefont {T.}~\bibnamefont {{Erben}}}, \bibinfo {author} {\bibfnamefont
  {R.}~\bibnamefont {{Gavazzi}}}, \bibinfo {author} {\bibfnamefont
  {S.}~\bibnamefont {{Gwyn}}},  \emph {et~al.},\ }\href {\doibase
  10.1051/0004-6361/202141847} {\bibfield  {journal} {\bibinfo  {journal}
  {\aap}\ }\textbf {\bibinfo {volume} {666}},\ \bibinfo {eid} {A162} (\bibinfo
  {year} {2022})},\ \Eprint {http://arxiv.org/abs/2204.04798}
  {arXiv:2204.04798} \BibitemShut {NoStop}%
\bibitem [{\citenamefont {{Li}}\ \emph {et~al.}(2024)\citenamefont {{Li}},
  \citenamefont {{Kilbinger}}, \citenamefont {{Luo}}, \citenamefont {{Wang}},
  \citenamefont {{Wang}}, \citenamefont {{Wittje}}, \citenamefont
  {{Hildebrandt}}, \citenamefont {{van Waerbeke}}, \citenamefont {{Hudson}},
  \citenamefont {{Farrens}} \emph {et~al.}}]{Li2024}%
  \BibitemOpen
  \bibfield  {author} {\bibinfo {author} {\bibfnamefont {Q.}~\bibnamefont
  {{Li}}}, \bibinfo {author} {\bibfnamefont {M.}~\bibnamefont {{Kilbinger}}},
  \bibinfo {author} {\bibfnamefont {W.}~\bibnamefont {{Luo}}}, \bibinfo
  {author} {\bibfnamefont {K.}~\bibnamefont {{Wang}}}, \bibinfo {author}
  {\bibfnamefont {H.}~\bibnamefont {{Wang}}}, \bibinfo {author} {\bibfnamefont
  {A.}~\bibnamefont {{Wittje}}}, \bibinfo {author} {\bibfnamefont
  {H.}~\bibnamefont {{Hildebrandt}}}, \bibinfo {author} {\bibfnamefont
  {L.}~\bibnamefont {{van Waerbeke}}}, \bibinfo {author} {\bibfnamefont
  {M.~J.}\ \bibnamefont {{Hudson}}}, \bibinfo {author} {\bibfnamefont
  {S.}~\bibnamefont {{Farrens}}},  \emph {et~al.},\ }\href@noop {} {\
  (\bibinfo {year} {2024})},\ \Eprint {http://arxiv.org/abs/2402.10740}
  {arXiv:2402.10740} \BibitemShut {NoStop}%
\bibitem [{\citenamefont {{Reid}}\ \emph {et~al.}(2016)\citenamefont {{Reid}},
  \citenamefont {{Ho}}, \citenamefont {{Padmanabhan}}, \citenamefont
  {{Percival}}, \citenamefont {{Tinker}}, \citenamefont {{Tojeiro}},
  \citenamefont {{White}}, \citenamefont {{Eisenstein}}, \citenamefont
  {{Maraston}}, \citenamefont {{Ross}} \emph {et~al.}}]{Reid2016}%
  \BibitemOpen
  \bibfield  {author} {\bibinfo {author} {\bibfnamefont {B.}~\bibnamefont
  {{Reid}}}, \bibinfo {author} {\bibfnamefont {S.}~\bibnamefont {{Ho}}},
  \bibinfo {author} {\bibfnamefont {N.}~\bibnamefont {{Padmanabhan}}}, \bibinfo
  {author} {\bibfnamefont {W.~J.}\ \bibnamefont {{Percival}}}, \bibinfo
  {author} {\bibfnamefont {J.}~\bibnamefont {{Tinker}}}, \bibinfo {author}
  {\bibfnamefont {R.}~\bibnamefont {{Tojeiro}}}, \bibinfo {author}
  {\bibfnamefont {M.}~\bibnamefont {{White}}}, \bibinfo {author} {\bibfnamefont
  {D.~J.}\ \bibnamefont {{Eisenstein}}}, \bibinfo {author} {\bibfnamefont
  {C.}~\bibnamefont {{Maraston}}}, \bibinfo {author} {\bibfnamefont {A.~J.}\
  \bibnamefont {{Ross}}},  \emph {et~al.},\ }\href {\doibase
  10.1093/mnras/stv2382} {\bibfield  {journal} {\bibinfo  {journal} {\mnras}\
  }\textbf {\bibinfo {volume} {455}},\ \bibinfo {pages} {1553} (\bibinfo {year}
  {2016})},\ \Eprint {http://arxiv.org/abs/1509.06529} {arXiv:1509.06529}
  \BibitemShut {NoStop}%
\bibitem [{\citenamefont {{Garrison}}\ \emph {et~al.}(2019)\citenamefont
  {{Garrison}}, \citenamefont {{Eisenstein}},\ and\ \citenamefont
  {{Pinto}}}]{Garrison2019}%
  \BibitemOpen
  \bibfield  {author} {\bibinfo {author} {\bibfnamefont {L.~H.}\ \bibnamefont
  {{Garrison}}}, \bibinfo {author} {\bibfnamefont {D.~J.}\ \bibnamefont
  {{Eisenstein}}}, \ and\ \bibinfo {author} {\bibfnamefont {P.~A.}\
  \bibnamefont {{Pinto}}},\ }\href {\doibase 10.1093/mnras/stz634} {\bibfield
  {journal} {\bibinfo  {journal} {\mnras}\ }\textbf {\bibinfo {volume} {485}},\
  \bibinfo {pages} {3370} (\bibinfo {year} {2019})},\ \Eprint
  {http://arxiv.org/abs/1810.02916} {arXiv:1810.02916} \BibitemShut {NoStop}%
\bibitem [{\citenamefont {{Garrison}}\ \emph {et~al.}(2021)\citenamefont
  {{Garrison}}, \citenamefont {{Eisenstein}}, \citenamefont {{Ferrer}},
  \citenamefont {{Maksimova}},\ and\ \citenamefont {{Pinto}}}]{Garrison2021}%
  \BibitemOpen
  \bibfield  {author} {\bibinfo {author} {\bibfnamefont {L.~H.}\ \bibnamefont
  {{Garrison}}}, \bibinfo {author} {\bibfnamefont {D.~J.}\ \bibnamefont
  {{Eisenstein}}}, \bibinfo {author} {\bibfnamefont {D.}~\bibnamefont
  {{Ferrer}}}, \bibinfo {author} {\bibfnamefont {N.~A.}\ \bibnamefont
  {{Maksimova}}}, \ and\ \bibinfo {author} {\bibfnamefont {P.~A.}\ \bibnamefont
  {{Pinto}}},\ }\href {\doibase 10.1093/mnras/stab2482} {\bibfield  {journal}
  {\bibinfo  {journal} {\mnras}\ }\textbf {\bibinfo {volume} {508}},\ \bibinfo
  {pages} {575} (\bibinfo {year} {2021})},\ \Eprint
  {http://arxiv.org/abs/2110.11392} {arXiv:2110.11392} \BibitemShut {NoStop}%
\bibitem [{\citenamefont {{Hadzhiyska}}\ \emph {et~al.}(2022)\citenamefont
  {{Hadzhiyska}}, \citenamefont {{Eisenstein}}, \citenamefont {{Bose}},
  \citenamefont {{Garrison}},\ and\ \citenamefont
  {{Maksimova}}}]{Hadzhiyska2022}%
  \BibitemOpen
  \bibfield  {author} {\bibinfo {author} {\bibfnamefont {B.}~\bibnamefont
  {{Hadzhiyska}}}, \bibinfo {author} {\bibfnamefont {D.}~\bibnamefont
  {{Eisenstein}}}, \bibinfo {author} {\bibfnamefont {S.}~\bibnamefont
  {{Bose}}}, \bibinfo {author} {\bibfnamefont {L.~H.}\ \bibnamefont
  {{Garrison}}}, \ and\ \bibinfo {author} {\bibfnamefont {N.}~\bibnamefont
  {{Maksimova}}},\ }\href {\doibase 10.1093/mnras/stab2980} {\bibfield
  {journal} {\bibinfo  {journal} {\mnras}\ }\textbf {\bibinfo {volume} {509}},\
  \bibinfo {pages} {501} (\bibinfo {year} {2022})},\ \Eprint
  {http://arxiv.org/abs/2110.11408} {arXiv:2110.11408} \BibitemShut {NoStop}%
\bibitem [{\citenamefont {{Planck Collaboration}}\ \emph
  {et~al.}(2020)\citenamefont {{Planck Collaboration}}, \citenamefont
  {{Aghanim}}, \citenamefont {{Akrami}}, \citenamefont {{Ashdown}},
  \citenamefont {{Aumont}}, \citenamefont {{Baccigalupi}}, \citenamefont
  {{Ballardini}}, \citenamefont {{Banday}}, \citenamefont {{Barreiro}},
  \citenamefont {{Bartolo}} \emph {et~al.}}]{Planck2020}%
  \BibitemOpen
  \bibfield  {author} {\bibinfo {author} {\bibnamefont {{Planck
  Collaboration}}}, \bibinfo {author} {\bibfnamefont {N.}~\bibnamefont
  {{Aghanim}}}, \bibinfo {author} {\bibfnamefont {Y.}~\bibnamefont {{Akrami}}},
  \bibinfo {author} {\bibfnamefont {M.}~\bibnamefont {{Ashdown}}}, \bibinfo
  {author} {\bibfnamefont {J.}~\bibnamefont {{Aumont}}}, \bibinfo {author}
  {\bibfnamefont {C.}~\bibnamefont {{Baccigalupi}}}, \bibinfo {author}
  {\bibfnamefont {M.}~\bibnamefont {{Ballardini}}}, \bibinfo {author}
  {\bibfnamefont {A.~J.}\ \bibnamefont {{Banday}}}, \bibinfo {author}
  {\bibfnamefont {R.~B.}\ \bibnamefont {{Barreiro}}}, \bibinfo {author}
  {\bibfnamefont {N.}~\bibnamefont {{Bartolo}}},  \emph {et~al.},\ }\href
  {\doibase 10.1051/0004-6361/201833910} {\bibfield  {journal} {\bibinfo
  {journal} {\aap}\ }\textbf {\bibinfo {volume} {641}},\ \bibinfo {eid} {A6}
  (\bibinfo {year} {2020})},\ \Eprint {http://arxiv.org/abs/1807.06209}
  {arXiv:1807.06209} \BibitemShut {NoStop}%
\bibitem [{\citenamefont {{R{\'a}cz}}\ \emph {et~al.}(2023)\citenamefont
  {{R{\'a}cz}}, \citenamefont {{Kiessling}}, \citenamefont {{Csabai}},\ and\
  \citenamefont {{Szapudi}}}]{Racz2023}%
  \BibitemOpen
  \bibfield  {author} {\bibinfo {author} {\bibfnamefont {G.}~\bibnamefont
  {{R{\'a}cz}}}, \bibinfo {author} {\bibfnamefont {A.}~\bibnamefont
  {{Kiessling}}}, \bibinfo {author} {\bibfnamefont {I.}~\bibnamefont
  {{Csabai}}}, \ and\ \bibinfo {author} {\bibfnamefont {I.}~\bibnamefont
  {{Szapudi}}},\ }\href {\doibase 10.1051/0004-6361/202245211} {\bibfield
  {journal} {\bibinfo  {journal} {\aap}\ }\textbf {\bibinfo {volume} {672}},\
  \bibinfo {eid} {A59} (\bibinfo {year} {2023})},\ \Eprint
  {http://arxiv.org/abs/2210.15077} {arXiv:2210.15077 [astro-ph.CO]}
  \BibitemShut {NoStop}%
\bibitem [{\citenamefont {{Zheng}}\ \emph {et~al.}(2007)\citenamefont
  {{Zheng}}, \citenamefont {{Coil}},\ and\ \citenamefont
  {{Zehavi}}}]{Zheng2007}%
  \BibitemOpen
  \bibfield  {author} {\bibinfo {author} {\bibfnamefont {Z.}~\bibnamefont
  {{Zheng}}}, \bibinfo {author} {\bibfnamefont {A.~L.}\ \bibnamefont {{Coil}}},
  \ and\ \bibinfo {author} {\bibfnamefont {I.}~\bibnamefont {{Zehavi}}},\
  }\href {\doibase 10.1086/521074} {\bibfield  {journal} {\bibinfo  {journal}
  {\apj}\ }\textbf {\bibinfo {volume} {667}},\ \bibinfo {pages} {760} (\bibinfo
  {year} {2007})},\ \Eprint {http://arxiv.org/abs/astro-ph/0703457}
  {arXiv:astro-ph/0703457 [astro-ph]} \BibitemShut {NoStop}%
\bibitem [{\citenamefont {{Yuan}}\ \emph {et~al.}(2022)\citenamefont {{Yuan}},
  \citenamefont {{Garrison}}, \citenamefont {{Hadzhiyska}}, \citenamefont
  {{Bose}},\ and\ \citenamefont {{Eisenstein}}}]{Yuan2022}%
  \BibitemOpen
  \bibfield  {author} {\bibinfo {author} {\bibfnamefont {S.}~\bibnamefont
  {{Yuan}}}, \bibinfo {author} {\bibfnamefont {L.~H.}\ \bibnamefont
  {{Garrison}}}, \bibinfo {author} {\bibfnamefont {B.}~\bibnamefont
  {{Hadzhiyska}}}, \bibinfo {author} {\bibfnamefont {S.}~\bibnamefont
  {{Bose}}}, \ and\ \bibinfo {author} {\bibfnamefont {D.~J.}\ \bibnamefont
  {{Eisenstein}}},\ }\href {\doibase 10.1093/mnras/stab3355} {\bibfield
  {journal} {\bibinfo  {journal} {\mnras}\ }\textbf {\bibinfo {volume} {510}},\
  \bibinfo {pages} {3301} (\bibinfo {year} {2022})},\ \Eprint
  {http://arxiv.org/abs/2110.11412} {arXiv:2110.11412} \BibitemShut {NoStop}%
\bibitem [{\citenamefont {{Yuan}}\ \emph {et~al.}(2018)\citenamefont {{Yuan}},
  \citenamefont {{Eisenstein}},\ and\ \citenamefont {{Garrison}}}]{Yuan2018}%
  \BibitemOpen
  \bibfield  {author} {\bibinfo {author} {\bibfnamefont {S.}~\bibnamefont
  {{Yuan}}}, \bibinfo {author} {\bibfnamefont {D.~J.}\ \bibnamefont
  {{Eisenstein}}}, \ and\ \bibinfo {author} {\bibfnamefont {L.~H.}\
  \bibnamefont {{Garrison}}},\ }\href {\doibase 10.1093/mnras/sty1089}
  {\bibfield  {journal} {\bibinfo  {journal} {\mnras}\ }\textbf {\bibinfo
  {volume} {478}},\ \bibinfo {pages} {2019} (\bibinfo {year} {2018})},\ \Eprint
  {http://arxiv.org/abs/1802.10115} {arXiv:1802.10115} \BibitemShut {NoStop}%
\bibitem [{\citenamefont {{Takahashi}}\ \emph {et~al.}(2017)\citenamefont
  {{Takahashi}}, \citenamefont {{Hamana}}, \citenamefont {{Shirasaki}},
  \citenamefont {{Namikawa}}, \citenamefont {{Nishimichi}}, \citenamefont
  {{Osato}},\ and\ \citenamefont {{Shiroyama}}}]{Takahashi2017}%
  \BibitemOpen
  \bibfield  {author} {\bibinfo {author} {\bibfnamefont {R.}~\bibnamefont
  {{Takahashi}}}, \bibinfo {author} {\bibfnamefont {T.}~\bibnamefont
  {{Hamana}}}, \bibinfo {author} {\bibfnamefont {M.}~\bibnamefont
  {{Shirasaki}}}, \bibinfo {author} {\bibfnamefont {T.}~\bibnamefont
  {{Namikawa}}}, \bibinfo {author} {\bibfnamefont {T.}~\bibnamefont
  {{Nishimichi}}}, \bibinfo {author} {\bibfnamefont {K.}~\bibnamefont
  {{Osato}}}, \ and\ \bibinfo {author} {\bibfnamefont {K.}~\bibnamefont
  {{Shiroyama}}},\ }\href {\doibase 10.3847/1538-4357/aa943d} {\bibfield
  {journal} {\bibinfo  {journal} {\apj}\ }\textbf {\bibinfo {volume} {850}},\
  \bibinfo {eid} {24} (\bibinfo {year} {2017})},\ \Eprint
  {http://arxiv.org/abs/1706.01472} {arXiv:1706.01472} \BibitemShut {NoStop}%
\bibitem [{\citenamefont {{Springel}}(2005)}]{Springel2005}%
  \BibitemOpen
  \bibfield  {author} {\bibinfo {author} {\bibfnamefont {V.}~\bibnamefont
  {{Springel}}},\ }\href {\doibase 10.1111/j.1365-2966.2005.09655.x} {\bibfield
   {journal} {\bibinfo  {journal} {\mnras}\ }\textbf {\bibinfo {volume}
  {364}},\ \bibinfo {pages} {1105} (\bibinfo {year} {2005})},\ \Eprint
  {http://arxiv.org/abs/astro-ph/0505010} {arXiv:astro-ph/0505010} \BibitemShut
  {NoStop}%
\bibitem [{\citenamefont {{Seitz}}\ and\ \citenamefont
  {{Schneider}}(1997)}]{Seitz1997}%
  \BibitemOpen
  \bibfield  {author} {\bibinfo {author} {\bibfnamefont {C.}~\bibnamefont
  {{Seitz}}}\ and\ \bibinfo {author} {\bibfnamefont {P.}~\bibnamefont
  {{Schneider}}},\ }\href@noop {} {\bibfield  {journal} {\bibinfo  {journal}
  {\aap}\ }\textbf {\bibinfo {volume} {318}},\ \bibinfo {pages} {687} (\bibinfo
  {year} {1997})},\ \Eprint {http://arxiv.org/abs/astro-ph/9601079}
  {arXiv:astro-ph/9601079} \BibitemShut {NoStop}%
\bibitem [{\citenamefont {{Harnois-D{\'e}raps}}\ \emph
  {et~al.}(2018)\citenamefont {{Harnois-D{\'e}raps}}, \citenamefont {{Amon}},
  \citenamefont {{Choi}}, \citenamefont {{Demchenko}}, \citenamefont
  {{Heymans}}, \citenamefont {{Kannawadi}}, \citenamefont {{Nakajima}},
  \citenamefont {{Sirks}}, \citenamefont {{van Waerbeke}}, \citenamefont
  {{Cai}} \emph {et~al.}}]{Harnois-Deraps:2018}%
  \BibitemOpen
  \bibfield  {author} {\bibinfo {author} {\bibfnamefont {J.}~\bibnamefont
  {{Harnois-D{\'e}raps}}}, \bibinfo {author} {\bibfnamefont {A.}~\bibnamefont
  {{Amon}}}, \bibinfo {author} {\bibfnamefont {A.}~\bibnamefont {{Choi}}},
  \bibinfo {author} {\bibfnamefont {V.}~\bibnamefont {{Demchenko}}}, \bibinfo
  {author} {\bibfnamefont {C.}~\bibnamefont {{Heymans}}}, \bibinfo {author}
  {\bibfnamefont {A.}~\bibnamefont {{Kannawadi}}}, \bibinfo {author}
  {\bibfnamefont {R.}~\bibnamefont {{Nakajima}}}, \bibinfo {author}
  {\bibfnamefont {E.}~\bibnamefont {{Sirks}}}, \bibinfo {author} {\bibfnamefont
  {L.}~\bibnamefont {{van Waerbeke}}}, \bibinfo {author} {\bibfnamefont
  {Y.-C.}\ \bibnamefont {{Cai}}},  \emph {et~al.},\ }\href {\doibase
  10.1093/mnras/sty2319} {\bibfield  {journal} {\bibinfo  {journal} {\mnras}\
  }\textbf {\bibinfo {volume} {481}},\ \bibinfo {pages} {1337} (\bibinfo {year}
  {2018})},\ \Eprint {http://arxiv.org/abs/1805.04511} {arXiv:1805.04511}
  \BibitemShut {NoStop}%
\bibitem [{\citenamefont {{Kobayashi}}\ \emph {et~al.}(2022)\citenamefont
  {{Kobayashi}}, \citenamefont {{Nishimichi}}, \citenamefont {{Takada}},\ and\
  \citenamefont {{Miyatake}}}]{Kobayashi2022}%
  \BibitemOpen
  \bibfield  {author} {\bibinfo {author} {\bibfnamefont {Y.}~\bibnamefont
  {{Kobayashi}}}, \bibinfo {author} {\bibfnamefont {T.}~\bibnamefont
  {{Nishimichi}}}, \bibinfo {author} {\bibfnamefont {M.}~\bibnamefont
  {{Takada}}}, \ and\ \bibinfo {author} {\bibfnamefont {H.}~\bibnamefont
  {{Miyatake}}},\ }\href {\doibase 10.1103/PhysRevD.105.083517} {\bibfield
  {journal} {\bibinfo  {journal} {\prd}\ }\textbf {\bibinfo {volume} {105}},\
  \bibinfo {eid} {083517} (\bibinfo {year} {2022})},\ \Eprint
  {http://arxiv.org/abs/2110.06969} {arXiv:2110.06969 [astro-ph.CO]}
  \BibitemShut {NoStop}%
\bibitem [{\citenamefont {{Jarvis}}\ \emph {et~al.}(2004)\citenamefont
  {{Jarvis}}, \citenamefont {{Bernstein}},\ and\ \citenamefont
  {{Jain}}}]{Jarvis:2004}%
  \BibitemOpen
  \bibfield  {author} {\bibinfo {author} {\bibfnamefont {M.}~\bibnamefont
  {{Jarvis}}}, \bibinfo {author} {\bibfnamefont {G.}~\bibnamefont
  {{Bernstein}}}, \ and\ \bibinfo {author} {\bibfnamefont {B.}~\bibnamefont
  {{Jain}}},\ }\href {\doibase 10.1111/j.1365-2966.2004.07926.x} {\bibfield
  {journal} {\bibinfo  {journal} {\mnras}\ }\textbf {\bibinfo {volume} {352}},\
  \bibinfo {pages} {338} (\bibinfo {year} {2004})},\ \Eprint
  {http://arxiv.org/abs/astro-ph/0307393} {arXiv:astro-ph/0307393} \BibitemShut
  {NoStop}%
\bibitem [{\citenamefont {{Aric{\`o}}}\ \emph {et~al.}(2021)\citenamefont
  {{Aric{\`o}}}, \citenamefont {{Angulo}}, \citenamefont {{Contreras}},
  \citenamefont {{Ondaro-Mallea}}, \citenamefont {{Pellejero-Iba{\~n}ez}},\
  and\ \citenamefont {{Zennaro}}}]{Arico2021}%
  \BibitemOpen
  \bibfield  {author} {\bibinfo {author} {\bibfnamefont {G.}~\bibnamefont
  {{Aric{\`o}}}}, \bibinfo {author} {\bibfnamefont {R.~E.}\ \bibnamefont
  {{Angulo}}}, \bibinfo {author} {\bibfnamefont {S.}~\bibnamefont
  {{Contreras}}}, \bibinfo {author} {\bibfnamefont {L.}~\bibnamefont
  {{Ondaro-Mallea}}}, \bibinfo {author} {\bibfnamefont {M.}~\bibnamefont
  {{Pellejero-Iba{\~n}ez}}}, \ and\ \bibinfo {author} {\bibfnamefont
  {M.}~\bibnamefont {{Zennaro}}},\ }\href {\doibase 10.1093/mnras/stab1911}
  {\bibfield  {journal} {\bibinfo  {journal} {\mnras}\ }\textbf {\bibinfo
  {volume} {506}},\ \bibinfo {pages} {4070} (\bibinfo {year} {2021})},\ \Eprint
  {http://arxiv.org/abs/2011.15018} {arXiv:2011.15018 [astro-ph.CO]}
  \BibitemShut {NoStop}%
\bibitem [{\citenamefont {{Percival}}\ \emph {et~al.}(2022)\citenamefont
  {{Percival}}, \citenamefont {{Friedrich}}, \citenamefont {{Sellentin}},\ and\
  \citenamefont {{Heavens}}}]{Percival2021}%
  \BibitemOpen
  \bibfield  {author} {\bibinfo {author} {\bibfnamefont {W.~J.}\ \bibnamefont
  {{Percival}}}, \bibinfo {author} {\bibfnamefont {O.}~\bibnamefont
  {{Friedrich}}}, \bibinfo {author} {\bibfnamefont {E.}~\bibnamefont
  {{Sellentin}}}, \ and\ \bibinfo {author} {\bibfnamefont {A.}~\bibnamefont
  {{Heavens}}},\ }\href {\doibase 10.1093/mnras/stab3540} {\bibfield  {journal}
  {\bibinfo  {journal} {\mnras}\ }\textbf {\bibinfo {volume} {510}},\ \bibinfo
  {pages} {3207} (\bibinfo {year} {2022})},\ \Eprint
  {http://arxiv.org/abs/2108.10402} {arXiv:2108.10402} \BibitemShut {NoStop}%
\bibitem [{\citenamefont {{Sellentin}}\ and\ \citenamefont
  {{Heavens}}(2016)}]{Sellentin2016}%
  \BibitemOpen
  \bibfield  {author} {\bibinfo {author} {\bibfnamefont {E.}~\bibnamefont
  {{Sellentin}}}\ and\ \bibinfo {author} {\bibfnamefont {A.~F.}\ \bibnamefont
  {{Heavens}}},\ }\href {\doibase 10.1093/mnrasl/slv190} {\bibfield  {journal}
  {\bibinfo  {journal} {\mnras}\ }\textbf {\bibinfo {volume} {456}},\ \bibinfo
  {pages} {L132} (\bibinfo {year} {2016})},\ \Eprint
  {http://arxiv.org/abs/1511.05969} {arXiv:1511.05969} \BibitemShut {NoStop}%
\bibitem [{\citenamefont {Shapiro}\ and\ \citenamefont {Wilk}(1965)}]{SHAPIRO}%
  \BibitemOpen
  \bibfield  {author} {\bibinfo {author} {\bibfnamefont {S.~S.}\ \bibnamefont
  {Shapiro}}\ and\ \bibinfo {author} {\bibfnamefont {M.~B.}\ \bibnamefont
  {Wilk}},\ }\href {\doibase 10.1093/biomet/52.3-4.591} {\bibfield  {journal}
  {\bibinfo  {journal} {Biometrika}\ }\textbf {\bibinfo {volume} {52}},\
  \bibinfo {pages} {591} (\bibinfo {year} {1965})},\ \Eprint
  {http://arxiv.org/abs/https://academic.oup.com/biomet/article-pdf/52/3-4/591/962907/52-3-4-591.pdf}
  {https://academic.oup.com/biomet/article-pdf/52/3-4/591/962907/52-3-4-591.pdf}
  \BibitemShut {NoStop}%
\end{thebibliography}%
 
\end{document}